%% file: main.tex
\documentclass[runningheads]{llncs}

\usepackage[utf8]{inputenc}
\usepackage[T1]{fontenc}
\usepackage[english]{babel}
\usepackage{listings}
\usepackage{enumerate}
\usepackage{hyperref}
\usepackage{color}

\usepackage{graphicx}
\usepackage{amsmath}
\usepackage{amssymb}

\usepackage{caption}
\usepackage{subcaption}
\usepackage{float}
\usepackage{todonotes}

\newtheorem{formalisation}{Formalisation}
\BeforeBeginEnvironment{formalisation}{\par\hangindent=0.3cm}
\AfterEndEnvironment{formalisation}{\par}

\input{commands}

\long\def\ignore#1{}

\begin{document}

\title{An Encoding of Abstract Dialectical Frameworks into Higher-Order Logic}
\author{Antoine Martina\inst{1} \and Alexander Steen\inst{2}\orcidID{0000-0001-8781-9462}}

\authorrunning{A. Martina and A. Steen}

\institute{}
\institute{University of Luxembourg, Department of Computer Science \\ 6, avenue de la Fonte \\ L-4364 Esch-sur-Alzette, Luxembourg \\
\and
Universität Greifswald, Institut für Mathematik und Informatik \\ Walther-Rathenau-Straße 47 \\ 17489 Greifswald, Germany \\
\email{alexander.steen@uni-greifswald.de}}
\maketitle

\begin{abstract}
  An approach for encoding abstract dialectical frameworks and their semantics into classical higher-order logic is presented.
  Important properties and semantic relationships are formally encoded and proven using the proof assistant Isabelle/HOL.
  This approach allows for the computer-assisted analysis of abstract dialectical frameworks using automated and interactive
  reasoning tools within a uniform logic environment. Exemplary applications include the formal analysis and verification of
  meta-theoretical properties, and the generation of interpretations and extensions under specific semantic constraints.
  
\end{abstract}

\section{Introduction}\label{sec:Intro}
Argumentation theory is an active field of study in artificial intelligence and non-monotonic reasoning.\ Argumentation frameworks (AFs) as introduced by Dung in 1995 \cite{AFsDung} have become a key concept in abstract argumentation, with applications to a variety of non-monotonic reasoning problems such as legal reasoning, logic programming, inconsistency handling and many others \cite{rahwan2009argumentation,zhang2009argumentation,amgoud2010handling,toni2011argumentation}. AFs are widely used in argumentation for handling conflicts among (abstract) arguments. One criticism against AFs is that only one type of relationship is allowed between arguments: an argument attacking another. Recently, abstract dialectical frameworks (ADFs) have been introduced \cite{bipolar_ADFs} as a powerful generalisation of Dung’s framework.
ADFs allow for arbitrary relationships between arguments whereas AFs only allow for attacks.
For example, ADFs allow for support relations between arguments, or relations where a particular group of arguments is required in order to successfully attack another specific argument.

Since Dung's original contribution, a lot of research has gone into algorithmic procedures, complexity aspects, and related and extended formalisms \cite{baroni2020acceptability}. A common method for implementing abstract argumentation is the encoding of constraints imposed by argumentation semantics into other formalisms \cite{cerutti2017foundations}. For example, the problem of finding extensions, i.e., sets of arguments that may be accepted together under a specific argumentation semantics, has been reduced to a logical satisfiability problem~\cite{besnard2004checking,dunne2002coherence}. This work has paved the way for works such as \cite{wallner2013advanced,cerutti2017foundations} where the power of SAT-solvers \cite{biere2009handbook} is harnessed to generate appropriate semantic extensions. This technique is also used as basis for various argumentation tools such as LabSATSolver \cite{beierle2015software}, jArgSemSAT \cite{cerutti2017efficient}, and Cegartix \cite{dvovrak2014complexity}. While these methods are quite efficient for computing extensions of large-scale AFs, they do not allow for meta-theoretical reasoning about abstract argumentation.

Beyond reductions to SAT, other approaches consist of encoding AFs resp.\ ADFs into more expressive logical formalisms than propositional logic \cite{egly2006reasoning,de2016argumentation,cayrol2020logical}. This allows to capture many different argumentation semantics on a purely logical basis, thus allowing for some form of automation regarding the verification of some semantic properties \cite{arieli2013qbf}.\footnote{See \cite{gabbay2013} and references therein for a more detailed survey and discussion about logical encodings, including second-order and modal approaches.} For ADFs, Diller et al.\ present the QADF system~\cite{DBLP:journals/argcom/DillerWW15} based on reductions to Quantified Boolean Formulas (QBF), and Ellmauthaler and Strass present the DIAMOND system~\cite{DBLP:conf/comma/EllmauthalerS14} based on reductions to Answer Set Programming (ASP).  Furthermore, Linsbichler et al.\ discuss advanced algorithms that make use of structural properties of the ADF at hand~\cite{DBLP:journals/ai/LinsbichlerMNWW22}. 
Recently, argumentation frameworks and their semantics have been encoded into extensional type theory~\cite{AFinHOL}, often simply referred to as classical higher-order logic (HOL). Like for other encodings, already existing automated reasoning tools can then be used for the computer-assisted assessment of AFs. A considerable advantage, however, of choosing HOL as target formalism is that it constitutes a uniform approach for both reasoning \emph{with} and \emph{about} AFs and their semantics, i.e., for combining object-level and meta-level reasoning.
Furthermore, the rich HOL language allows the instantiation of abstract arguments with arbitrary complex structures, and also for reasoning about (the existence of) extensions satisfying arbitrary higher-order properties.

In this article the HOL encoding of AFs of Steen and Fuenmayor~\cite{AFinHOL} is extended and generalised to ADFs and their semantics, bridging the gap between ADFs and automated reasoning in order to provide generic means to automatically and interactively assess ADFs within a rich ecosystem of higher-order reasoning tools, such as automated theorem provers and proof assistants.

Automated theorem provers (ATPs) are computer programs that attempt to prove a given conjecture from a given a set of assumptions. These proofs are found autonomously, meaning without any feedback or interaction from/with the user. On the other hand, interactive theorem provers  (also called proof assistants) are computer programs that assess the correctness of formal proofs created and given as input by the user. Proof assistants are usually based on (extensions of) HOL, cf. Sect.~\ref{sec:HOL}.
In this article, the well-established proof assistant called Isabelle/HOL \cite{Isabelle/HOL} is exemplarily employed.

The article is structured as follows. In \S 2 various preliminaries on ADFs, approximation fixpoint-theory and HOL are briefly introduced. 
\S 3 describes the encoding of ADFs and their semantics into HOL
In \S 4 important meta-theoretical properties are verified using Isabelle/HOL, and two application examples are presented.
Finally, \S 5 concludes and sketches further work.

\section{Preliminaries}
This section introduces ADFs and their semantics through a brief exposition of approximation fixpoint-theory (AFT).
Subsequently, a brief summary of the syntax and semantics of HOL is presented.
In this article, the latter formalism is used as expressive host language for encoding the former.

It should be noted that throughout this article different symbols are used for concepts on different (meta-)language levels: Set theoretic functions
are denoted with $\longrightarrow$ and $\mapsto$, e.g., $f:M \to N$ denotes 
a (set-theoretic) function from (set theoretic) domain $M$ to co-domain $N$ with $x \mapsto f(x)$ for every $x \in M$.
Concepts of object-language HOL are denoted with double-stroke arrows:
The function type constructor of HOL is denoted $\ar$ so that $f_{\tau \ar \nu}$ is a term of functional type $\tau \ar \nu$ of HOL (see details in Sect.~\ref{sec:HOL} below). Similarly, $\Longrightarrow$ denotes material implication in HOL.

\subsection{Formal Argumentation}
ADFs~\cite{ADFs} are generalisations of AFs~\cite{AFsDung,AFsHandbook}.
In essence, AFs are directed graphs; their nodes represent abstract arguments in the sense that the arguments themselves do not carry any specific meaning or structure, and the directed edges of the graph represent attacks between arguments. 
The goal is to identify a family of subsets of arguments, each of which may be considered jointly acceptable, with the remaining arguments deemed rejected or incompatible. Argumentation semantics formalize this concept, and the subsets of accepted arguments are referred to as extensions.

The limitations and AFs and the motivation of ADFs are thoroughly discussed in the literature~\cite{ADFs}.
Intuitively, in AFs, an argument $a$ attacked by two other arguments $b$ and $c$ is accepted iff both $b$ and $c$ are not accepted (though details can vary, depending on the argumentation semantics employed). This can be reformulated as $a$ having as acceptance condition the propositional formula $\neg b \land \neg c$. Generalising this to any AF, an argument's acceptance condition is then constructed by taking the conjunction of the negation of all its attackers. In AFs, these acceptance conditions are typically kept implicit since they are all of the same type. In ADFs however, acceptance conditions are made explicit as above but with the crucial difference that these conditions may be arbitrary propositional formulas.

For the rest of this section, material from \cite{ADFs} is adapted to introduce ADFs and their semantics.

\subsubsection{Approximation Fixpoint Theory.}\label{sec:AFT_and_AFs}
In the approximation fixpoint theory (AFT) framework~\cite{DMT2000}, knowledge bases are associated with approximation operators acting on algebraic structures. The fixpoints of these operators are then studied in order to analyse the semantics of knowledge bases.

In classical approaches to fixpoint-based semantics, the underlying algebraic structure is the complete lattice of the set $V_2 = \{v : A \longrightarrow \{\textbf{t}, \textbf{f}\}\}$ of all two-valued interpretations over some vocabulary $A$ (with \textbf{t} and \textbf{f}\ the Boolean truth values) ordered by the truth ordering $\leq_t$:
\begin{center}
    $v_1 \leq_t v_2$ \ iff \ $\forall a \in A : v_1(a) = \textbf{t} \implies v_2(a) = \textbf{t}$.
\end{center}
Furthermore, an operator $O$ on this lattice $(V_2, \leq_t)$ takes as input a two-valued interpretation $v \in V_2$ and returns a revised interpretation $O(v) \in V_2$. The intuition of the operator is that the revised interpretation $O(v)$ incorporates additional knowledge that is induced by the knowledge base associated to $O$ from interpretation $v$. Based on this intuition, fixpoints of $O$ correspond to the models of the knowledge base.

In AFT, the study of fixpoints of an operator $O$ is reduced to investigating fixpoints of its \textit{approximating operator} $\mathcal{O}$~\cite{DMT2000}. Whereas $O$ operates on two-valued interpretations over some vocabulary $A$, its approximation $\mathcal{O}$ operates on three-valued interpretations over $A$ represented by the set $V_3 = \{v : A \longrightarrow \{\textbf{t}, \textbf{f}, \textbf{u}\}\}$ (with the truth values \textbf{t} (true), \textbf{f} (false), and \textbf{u} (undefined)) and ordered by the information ordering $\leq_i$:
\begin{center}
    $v_1 \leq_i v_2$ \ iff \ $\forall a \in A: v_1(a) \in \{\textbf{t}, \textbf{f}\} \implies v_2(a) = v_1(a)$.
\end{center}
Intuitively, this ordering assigns a greater information content to the classical truth values $\{\textbf{t}, \textbf{f}\}$ than to undefined \textbf{u}. Furthermore, a single three-valued interpretation $v : A \longrightarrow \{\textbf{t}, \textbf{f}, \textbf{u}\}$ serves to approximate a set of two-valued interpretations
\begin{center}
    $[v]_2 := \{w \in V_2\ |\ v \leq_i w\}$.
\end{center}
For example, for the vocabulary $A = \{a, b, c\}$, the three-valued interpretation $v = \{a \mapsto \textbf{t}, b \mapsto \textbf{u}, c \mapsto \textbf{f}\}$ approximates the set $[v]_2 = \{w_1, w_2\}$ of two-valued interpretations where $w_1 = \{a \mapsto \textbf{t}, b \mapsto \textbf{t}, c \mapsto \textbf{f}\}$ and $w_2 = \{a \mapsto \textbf{t}, b \mapsto \textbf{f}, c \mapsto \textbf{f}\}$.

Similarly, a three-valued operator $\mathcal{O} : V_3 \longrightarrow V_3$ is said to approximate a two-valued operator $O : V_2 \longrightarrow V_2$ iff
\begin{enumerate}
    \item $\forall v \in V_2 \subseteq V_3$, $\mathcal{O}(v) = O(v)$ \ \ \ \ \ ($\mathcal{O}$ agrees with $O$ on two-valued $v$),
    \item $\forall v_1, v_2 \in V_3$, $v_1 \leq_i v_2 \implies \mathcal{O}(v_1) \leq_i \mathcal{O}(v_2)$ \ \ \ \ \ ($\mathcal{O}$ is $\leq_i$-monotone).
\end{enumerate}
\noindent In this case, fixpoints of $\mathcal{O}$ approximate fixpoints of $O$ in the sense that, for every fixpoint $w$ of $O$, there exists a fixpoint $v$ of $\mathcal{O}$ such that $w \in [v]_2$~\cite{DMT2000}. Moreover, an approximating operator $\mathcal{O}$ always has a fixpoint, which need not be the case for two-valued operators $O$. In particular, $\mathcal{O}$ has a $\leq_i$-least fixpoint, which approximates all fixpoints of $O$.

In subsequent work \cite{DMT2004}, a general and abstract way to define the most precise approximation of a given operator $O : V_2 \longrightarrow V_2$ is presented. Most precise here refers to a generalisation of $\leq_i$ to operators, where for $\mathcal{O}_1, \mathcal{O}_2 : V_3 \longrightarrow V_3$, they define $\mathcal{O}_1 \leq_i \mathcal{O}_2$ iff for all $v \in V_3$ it holds that $\mathcal{O}_1(v) \leq_i \mathcal{O}_2(v)$. Specifically, it is shown that the most precise, denoted \emph{ultimate}, approximation of $O$ is given by the operator $\mathcal{U}_O : V_3 \longrightarrow V_3$ that maps a given $v \in V_3$ to 
\begin{center}
    $\mathcal{U}_O(v) : A \longrightarrow \{\textbf{t}, \textbf{f}, \textbf{u}\}$ \ with \ $a \mapsto \left\{ 
    \begin{array}{l l}
        \textbf{t} & \text{if} \ w(a) = \textbf{t} \ \ \text{for all} \ \ w \in O\big([v]_2\big) = \{O(x)\ |\ x \in [v]_2\}\\
        \textbf{f} & \text{if} \ w(a) = \textbf{f} \ \ \text{for all} \ \ w \in O\big([v]_2\big) = \{O(x)\ |\ x \in [v]_2\}\\
        \textbf{u} & \text{otherwise}\\
    \end{array}
    \right.$
\end{center}
Note that, using this definition, approximations of operators can be derived uniformly.


\subsubsection{Abstract Dialectical Frameworks.}\label{sec:ADFs}
Like AFs, an ADF is a directed graph whose nodes represent arguments, statements or positions. The links (edges of the graph) represent dependencies. However, unlike a link in an AF, the meaning of an ADF link can vary. The status of a node $s$ only depends on the status of its parents (denoted $par(s)$), that is the nodes with a directed link towards $s$. In addition, each node $s$ has an associated acceptance condition $C_s$ specifying the conditions under which $s$ is accepted. $C_s$ is a function assigning to each subset of $par(s)$ one of the truth values $\textbf{t}, \textbf{f}$. Intuitively, if for some $R \subseteq par(s)$ it holds that $C_s(R) = \textbf{t}$, then $s$ will be accepted provided the nodes in $R$ are accepted and those in $par(s) \setminus R$ are not accepted.
\begin{definition}[Abstract Dialectical Framework]\label{def:ADF}
An \textit{abstract dialectical framework} $D$ is a tuple $D = (S,L,C)$ where:
\begin{itemize}
    \item $S$ is a set of statements (nodes),
    \item $L \subseteq S \times S$ is a set of links,
    \item $C = \{C_s\}_{s\in S}$ is a set of total functions $C_s : 2^{par(s)} \longrightarrow \{\textbf{t}, \textbf{f}\}$, where $2^{par(s)}$ denotes the power set of the set of parents of $s$. $C_s$ is called the \textit{acceptance condition} of statement $s$.
\end{itemize}
\end{definition}

\noindent In many cases, it is more practical and convenient to represent acceptance conditions as propositional formulas. For this reason, a slightly adapted representation of ADFs is employed for the rest of this preliminary section: an ADF $D = (S,L,C)$ is a tuple as above but with each acceptance condition $C_s$ represented as a propositional formula $\varphi_s$ over the vocabulary $par(s)$, i.e.\ $C = \{\varphi_s\}_{s\in S}$. 

Essentially, Def.~\ref{def:ADF} with acceptance conditions $C_s$ corresponds to the explicit description of the truth tables of their associated formulas $\varphi_s$. 

\begin{remark}
Given a two-valued interpretation $v : S \longrightarrow \{\textbf{t}, \textbf{f}\}$ and an acceptance condition $\varphi_s$ (i.e.\ a propositional formula built on the vocabulary $par(s) \subseteq S$), $v(\varphi_s)$ denotes the Boolean evaluation of formula $\varphi_s$. 
For example, if $v = \{a \mapsto \textbf{t}, b \mapsto \textbf{f}\}$, $\varphi_a = a \land b$ and $\varphi_b = a \lor b$, then $v(\varphi_a) = \textbf{f}$ and $v(\varphi_b) = \textbf{t}$.
\end{remark}

\begin{definition}[Two-valued Model]
A \textit{two-valued model} of an ADF is a two-valued interpretation $v : S \longrightarrow \{\textbf{t}, \textbf{f}\}$ such that for every statement $s \in S$, $v(s) = v(\varphi_s)$.
\end{definition}

\begin{example}\label{ex:ADF}
The ADF presented in Fig.~\ref{fig:exampleADF} acts as running example throughout this preliminary section.
The graph structure visualisation corresponds to the following specification in terms of Def.~\ref{def:ADF}:

\begin{figure}[tb]
    \centering
    \includegraphics[width=0.35\textwidth]{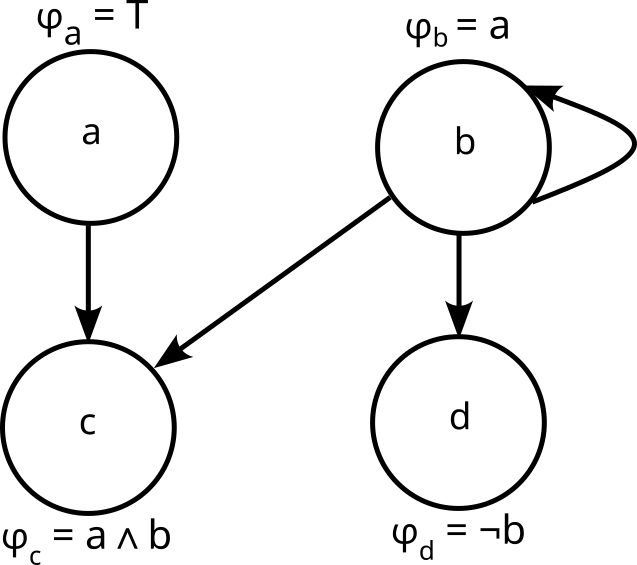}
    \caption{Visualisation of an example ADF taken from~\cite[Example 3.2]{ADFs}.}
    \label{fig:exampleADF}
\end{figure}

\begin{itemize}
    \item $S = \{a, b, c, d\}$
    \item $L = \{(a,c), (b,b), (b,c), (b,d)\}$
    \item $par(a) = \emptyset $ \ and \ $C_a = \{\emptyset \mapsto \textbf{t}\}$
    \item $par(b) = \{b\} $ \ and \ $C_b = \{\{b\} \mapsto \textbf{t}, \ \emptyset \mapsto \textbf{f}\}$
    \item $par(c) = \{a, b\} $ \ and \ $C_c = \{\{a, b\} \mapsto \textbf{t}, \ \{a\} \mapsto \textbf{f}, \ \{b\} \mapsto \textbf{f}, \ \emptyset \mapsto \textbf{f}\}$
    \item $par(d) = \{b\} $ \ and \ $C_d = \{\{b\} \mapsto \textbf{f}, \ \emptyset \mapsto \textbf{t}\}$
\end{itemize}
\noindent Furthermore, the above ADF has two two-valued models, namely $v_1 = \{a \mapsto  \textbf{t}, b \mapsto \textbf{t}, c \mapsto \textbf{t}, d \mapsto \textbf{f}\}$ and $v_2 = \{a \mapsto  \textbf{t}, b \mapsto \textbf{f}, c \mapsto \textbf{f}, d \mapsto \textbf{t}\}$.
\end{example}

\paragraph{Argumentation Semantics.}
Argumentation semantics for ADFs~\cite{Strass2013a,Brewka2013} are now defined in terms of ADF models and AFT approximation operators.
\begin{definition}[Two-valued Model Operator]\label{def:G_operator}
Let $D = (S,L,C)$ be an ADF, with $C = \{\varphi_s\}_{s \in S}$. The operator $G_D : V_2 \longrightarrow V_2$ takes as input a two-valued interpretation $v : S \longrightarrow \{\textbf{t}, \textbf{f}\}$ and returns an updated interpretation
\begin{center}
$G_D(v) : S \longrightarrow \{\textbf{t}, \textbf{f}\}$ \ with \ $s \mapsto v(\varphi_s)$
\end{center}
\end{definition}

\noindent
The fixpoints of this operator precisely correspond to the two-valued models of $D$ \cite{Strass2013a}.
It is then straightforward to determine the ultimate approximation of $G_D$ \cite{Strass2013a}:
\begin{definition}[Ultimate Approximation Operator]\label{def:Gamma_operator}
Let $D$ be an ADF. The operator $\Gamma_D : V_3 \longrightarrow V_3$ is the \emph{ultimate approximation of $G_D$}, given by
\begin{center}
    $\Gamma_D(v) : S \longrightarrow \{\textbf{t}, \textbf{f}, \textbf{u}\}$ \ with \ $s \mapsto \sqcap_i \{w(\varphi_s)\ |\ w \in [v]_2\}$,
\end{center}
where $v$ is a three-valued interpretation $v$, and $\sqcap_i$ is a "consensus" operation which assigns $\textbf{t} \sqcap_i \textbf{t} = \textbf{t}$, $\textbf{f} \sqcap_i \textbf{f} = \textbf{f}$, and returns \textbf{u} otherwise.
\end{definition}

\noindent Intuitively, for each statement $s$, the revised interpretation $\Gamma_D(v)$ returns the consensus truth value for the acceptance formula $\varphi_s$, where the consensus takes into account all possible two-valued interpretations $w$ that are approximated by the input three-valued interpretation $v$. If this $v$ is two-valued, then $[v]_2 = \{v\}$, thus $\Gamma_D(v)(s) = v(\varphi_s) = G_D(v)(s)$ and $\Gamma_D$ indeed approximates $G_D$.

\begin{definition}[Admissible, Complete, Preferred and Grounded Semantics]
Let $D = (S,L,C)$ be an ADF and $v : S \longrightarrow \{\textbf{t}, \textbf{f}, \textbf{u}\}$ be a three-valued interpretation.
\begin{enumerate}
    \item $v$ is \textit{admissible} for $D$ iff $v \leq_i \Gamma_D(v)$.
    \item $v$ is \textit{complete} for $D$ iff $v = \Gamma_D(v)$.
    \item $v$ is \textit{preferred} for $D$ iff $v$ is $\leq_i$-maximal admissible.
    \item $v$ is \textit{grounded} for $D$ iff $v$ is the $\leq_i$-least fixpoint of $\Gamma_D$.
\end{enumerate}
\end{definition}

\noindent Intuitively, a three-valued interpretation is admissible for an ADF $D$ iff it does not make an unjustified commitment that the operator $\Gamma_D$ will subsequently revoke. Additionally, the grounded semantics can be seen as the greatest possible consensus between all acceptable ways of interpreting the ADF at hand.

\begin{example}
Considering again the ADF $D$ from Example \ref{ex:ADF}, the following to be two-valued models exist:
\begin{itemize}
    \item $v_1 = \{a \mapsto  \textbf{t}, b \mapsto \textbf{t}, c \mapsto \textbf{t}, d \mapsto \textbf{f}\}$
    \item $v_2 = \{a \mapsto  \textbf{t}, b \mapsto \textbf{f}, c \mapsto \textbf{f}, d \mapsto \textbf{t}\}$
\end{itemize}
The interpretation $v_3 = \{a \mapsto  \textbf{t}, b \mapsto \textbf{u}, c \mapsto \textbf{u}, d \mapsto \textbf{u}\}$ is admissible since $\Gamma_D(v_3)(a) = v_3(a) = \textbf{t}$, implying $v_3 \leq_i \Gamma_D(v_3)$. On the other hand, the interpretation $v_4 = \{a \mapsto  \textbf{f}, b \mapsto \textbf{u}, c \mapsto \textbf{u}, d \mapsto \textbf{u}\}$ is not admissible since $v_4(a) = \textbf{f} \neq \textbf{t} = \Gamma_D(v_4)(a)$, implying $v_4 \nleq_i \Gamma_D(v_4)$.

Additionally, it can be verified that the complete interpretations of the example ADF $D$ are $v_1$, $v_2$ and $v_3$, with the first two also corresponding to the preferred interpretations. Finally, the grounded interpretation of $D$ is $v_3$.
\end{example}

The definition of stable semantics is taken from~\cite{Brewka2013}, cf.~\cite{Strass&Wallner2015} for an alternative definition via AFT.
\begin{definition}[Stable Model Semantics]\label{def:stable_semantics}
Let $D = (S,L,C)$ be an ADF with $C = \{\varphi_s\}_{s \in S}$ and $v : S \longrightarrow \{\textbf{t}, \textbf{f}\}$ be a two-valued model of $D$. The reduced ADF $D^v$ (or \textit{reduct}) is defined by $D^v = (S^v,L^v,C^v)$, where \ldots
\begin{itemize}
    \item $S^v = \{s \in S\ |\ v(s) = \textbf{t}\}$,
    \item $L^v = L \cap (S^v \times S^v$), and
    \item $C^v = \{\varphi_s^v\}_{s \in S^v}$ where for each $s \in S^v$, $\varphi_s^v := \varphi_s$ but with each node $a$ such that $v(a) = \textbf{f}$ replaced by $\bot$.
\end{itemize}
Now let $w$ be the unique grounded interpretation of $D^v$. The two-valued model $v$ is a \textit{stable model} of $D$ iff for all $s \in S$ it holds that that $v(s) = \textbf{t}$ implies $w(s) = \textbf{t}$.
\end{definition}

\noindent The basic intuition of stable semantics is that all elements of a stable model have a non-cyclic justification.

\begin{example}
Continuing the previous example ADF, it holds that
the grounded interpretation of reduct $D^{v_1}$ is $\{a \mapsto \textbf{t}, b \mapsto \textbf{u}, c \mapsto \textbf{u}\}$. $v_1$ is thus not a stable model of $D$ (note the cyclic justification of statement b). For $v_2$, the reduct $D^{v_2}$ has the grounded interpretation $\{a \mapsto \textbf{t}, d \mapsto \textbf{t}\}$. The model $v_2$ of $D$ is thus the single stable model of $D$.
\end{example}

\noindent Well-known relationships between semantics defined on Dung's AFs carry over to ADFs~\cite{Brewka2013}:
\begin{theorem}[Relations between ADF semantics]\label{theorem:relation_between_ADF_semantics}
Let $D$ be an ADF.
\begin{itemize}
    \item Each stable model of $D$ is a two-valued model of $D$;
    \item each two-valued model of $D$ is a preferred interpretation of $D$;
    \item each preferred interpretation of $D$ is complete;
    \item each complete interpretation of $D$ is admissible;
    \item the grounded interpretation of $D$ is complete.
\end{itemize}
\end{theorem}

\begin{theorem}[Existence of interpretations]\label{theorem:existance_semantics}
Let $D$ be an ADF. Then $D$ has at least one admissible, one complete, one preferred and one grounded interpretation. 
\end{theorem}


\subsection{Higher-Order Logic} \label{sec:HOL}
The following brief introduction to HOL is adopted from earlier work~\cite{AFinHOL}.

The term \textit{higher-order logic} refers to expressive logical formalisms that allow for quantification over predicate and function variables.
In the context of automated reasoning, higher-order logic commonly refers to systems
based on a simply typed $\lambda$-calculus, as originally introduced in the works of Church, Henkin and several others~\cite{Church,Henkin1}.
In the present work, higher-order logic (abbreviated as HOL) is used interchangeably with Henkin's Extensional Type Theory, cf.~\cite[\S 2]{AlexPhD}, which constitutes the basis of most contemporary higher-order automated reasoning systems.
HOL provides $\lambda$-notation as an expressive binding mechanism to denote unnamed functions, predicates and sets (by their characteristic functions), and it comes with built-in principles of Boolean and functional extensionality as well as type-restricted comprehension (cf.\ further below).

\paragraph{Syntax and Semantics.}
HOL is a typed logic; and all terms of HOL get assigned a fixed and unique type.
The set $\T$ of types is freely generated from a set of
base types $\B\T$ and the function type constructor $\ar$ (written as a right-associative infix operator).
Traditionally, the generating set $\B\T$ is taken to include at least two base types, $\B\T \subseteq \{\iota, \bo\}$,
where $\iota$ is interpreted as the type of individuals and $\bo$ as the type of (bivalent) Boolean truth values.

HOL terms of are given by the following abstract syntax
($\tau,\nu \in \T$):
\begin{equation*}
s,t ::= c_\tau \in \Sigma \; | \; X_\tau \in \V \; | \; \left(\lambda X_\tau.\, s_\nu\right)_{\tau\ar\nu}
  \; | \; \left(s_{\tau\ar\nu} \; t_\tau\right)_\nu
\end{equation*}
where $\Sigma$ is a set of constant symbols and $\V$ a set of variable symbols. The different forms of terms above are called \textit{constants}, \textit{variables}, \textit{abstractions} and \textit{applications}, respectively.
It is assumed that $\Sigma$ contains
equality predicate symbols $=^\tau_{\tau\ar\tau\ar \bo}$ for each
$\tau \in \T$. All remaining logical connectives can
be defined as abbreviations using equality and the other syntactical
structures~\cite[\S 2.1]{AlexPhD}, including
conjunction $\land_{\bo\ar\bo\ar\bo}$, disjunction $\lor_{\bo\ar\bo\ar\bo}$, 
material implication $\Longrightarrow_{\bo\ar\bo\ar\bo}$,
negation $\neg_{\bo\ar\bo}$, universal quantification for predicates
over type $\tau$ denoted $\Pi^\tau_{(\tau\ar\bo)\ar\bo}$.

For simplicity, the binary logical connectives may be written in infix notation,
e.g., the term/formula $p_o \lor q_o$ formally represents the application
$\left(\lor_{\bo\ar\bo\ar\bo} \; p_o \; q_o\right)$.
Also, so-called \emph{binder notation}~\cite{Currying} is used for universal and existential
quantification: The term $\forall X_\tau.\, s_o$ is used as a short-hand for
$\Pi^\tau_{(\tau\ar\bo)\ar\bo} \left(\lambda X_\tau.\, s_o \right)$
and analogously for existential quantification $\exists X_\tau.\, s_o$.
To improve readability, type-subscripts and parentheses are usually omitted
if there is no risk of confusion. 
Note that, by construction, HOL syntax only admits functions that take one parameter;
$n$-ary function applications are represented using \emph{currying}~\cite{Currying}, e.g.,
a first-order-like term such as $f(a,b)$ involving a binary function $f$ and two constants $a$ and
$b$ is represented in HOL
by consecutive applications of the individual constants,
as in $((f_{\iota\ar\iota\ar\iota}\; a_\iota)\; b_\iota)$,
or simply $f\;a\;b$ if omitting parentheses and type subscripts.
Here, the term $(f\; a)$ itself represents a function that is subsequently applied
to the argument $b$.
Also, functional terms may be only \emph{partially applied}, e.g., occurring in terms
like $(g_{(\iota\ar\iota)\ar\iota}\;(f\; a))$, where $f$ is the ``binary'' function from above
and $g_{(\iota\ar\iota)\ar\iota}$ is a higher-order function symbol taking a functional
expression of type $\iota\ar\iota$ as argument.

HOL automation is usually investigated with respect to so-called
\emph{general semantics}, due to Henkin~\cite{Henkin1},
for which complete proof calculi can be achieved.
Note that standard models for HOL are subsumed by general models
such that every valid formula with respect to general semantics is also
valid in the standard sense.
A formal exposition to HOL semantics is available in 
the literature (cf., e.g.,~\cite{AlexPhD,Henkin1}
and the references therein). For the remainder of this article,
HOL with general semantics is assumed.

\paragraph{HOL automation.}
There exist various interactive and automated theorem proving systems for HOL and variants thereof.
Isabelle/HOL~\cite{Isabelle/HOL} is an interactive theorem proving system that is widely employed for formalization tasks.
Isabelle/HOL implements Church's Simple Type Theory, i.e., a variant of HOL that is achieved
by enriching ExTT (as introduced above) with the axioms of infinity and description.\footnote{Note that none of the results
presented in this work depend on these additional assumptions.}
Isabelle/HOL itself is based on the general Isabelle system that uses a small trusted logic inference kernel, and thus
all (possibly large) proof objects of Isabelle/HOL are internally reduced
to inferences within that trusted core.
In principle, proofs are constructed by human users, e.g.\ in the Isabelle/Isar proof format~\cite{DBLP:conf/tphol/WenzelP06},
but can be assisted and shortened by internally verified proof tactics and proof automation procedures.
A significant practical feature of Isabelle/HOL is the 
Sledgehammer system~\cite{Sledgehammer} that bridges
between the proof assistant and external ATP systems, 
such as the first-order ATP system E~\cite{E-prover} 
or the higher-order ATP system Leo-III~\cite{SB2021}, and SMT solvers
such as Z3~\cite{Z3-solver} and CVC4~\cite{CVC4-solver}. 
The idea is to use these systems to automatically resolve
open proof obligations from Isabelle/HOL and to import and reconstruct the resulting (untrusted) proofs from the
external systems into
the verified context of Isabelle/HOL.
Additionally, Isabelle/HOL integrates so-called \emph{model finders}
such as Nitpick~\cite{Nitpick} that can generate (counter-)models to given formulas.
Specifically, Nitpick is a system for generating finite higher-order models; this is done
by a systematic enumeration and subsequent evaluation of possible model structures. The implementation makes use
of different heuristics and optimizations, and can regularly provide comprehsive counter-examples
within seconds in practice. Of course, with growing complexity of the logical the 

An extensive survey on HOL automation can be found in the literature~\cite{automationOfHOL,Currying}.

\section{Encoding of ADFs in HOL}\label{sec:ADFs_in_HOL}
The encoding of ADFs in HOL is now presented. The encoding's source files for Isabelle/HOL are freely available at Zenodo~\cite{files}.
Following most definitions, simple or well-known properties are verified by Isabelle/HOL.\footnote{
It should noted that ''verified  by Isabelle/HOL'' refers to the fact that formal and internally verified proofs have been automatically constructed by several theorem provers integrated into Isabelle/HOL via Sledgehammer. Furthermore, counter-examples to properties that indeed should not hold are generated by the Nitpick model-finder integrated into Isabelle/HOL.
} 
Each stated property is given the same name in the formalization artifacts, i.e., the corresponding Isabelle/HOL theory files \cite{files}.\footnote{The theory files have been developed using Isabelle 2021, the theory files have also been tested in combination with the more recent versions Isabelle 2022 and Isabelle 2023.}

In the definitions below \textit{type variables}, representing fixed but arbitrary types, are frequently used. This is motivated by the fact that  ADF statements (the nodes of an ADF graph) are not a priori associated with any specific structure. Depending on the context of the application in which the ADF formalism is being used, statements may represent, for example, actual natural language sentences just as much as propositional formulas or any other relevant object.
Following Isabelle/HOL notation conventions, type variables are represented using letters preceded by a single quote; e.g. $'s$ is a type variable. Additionally, throughout the following discussion, generous use of \textit{definitions}, as understood in the context of Isabelle/HOL is used. A definition defines a new symbol that, for the purposes of this article, is simply regarded as an abbreviation of some HOL term. For example, $a := b$ denotes the definition of a new symbol $a$ that stands for the (more complex) HOL term $b$. Finally and similarly to definitions, \textit{type synonyms} introduce a new type symbol that abbreviates a (more complex) type expression.

\subsection{Key Ideas and Basic Concepts}
In HOL a set is typically represented by its characteristic function, i.e.\ the function that maps to $true$ only the elements that belong in the set, and maps to $false$ otherwise. The type synonym $'s\ \texttt{Set}$ abbreviates the type $'s \ar o$. Similarly, relations are represented through their characteristic function as well.\footnote{Isabelle/HOL comes with a standard library that already provides definitions of sets, relations, and many other common mathematical structures, as well as proofs for many relevant properties of them. In this work, these axiomatically defined built-in types of Isabelle/HOL are not used. This is motivated by (1) the direct correspondence between the presented encoding and its representation in Isabelle/HOL, and (2) to underline the fact that this work does not depend on the Isabelle/HOL ecosystem in any way -- other theorem provers could be used instead.} 
The type synonym $'s\ \texttt{Rel}$ abbreviates the type $'s\ \ar\ 's\ \ar\ o$ associated with binary relations on terms of type $'s$. Finally, the type synonym $'s\ \texttt{Cond}$ abbreviates the type $'s\ \ar\ 's\ \texttt{Set}\ \ar\ o$, associated with the set of acceptance conditions $C$. Indeed, $C$ is a function that takes as input a statement $s$ of type $'s$ and returns a function $C_s$ which maps each subset of the parents of $s$ (sets of type $'s\ \texttt{Set}$) to one of the Boolean truth values \textbf{t} or \textbf{f}. The most important types and type synonyms are summarized in Table~\ref{tab:typeSynonyms}.

\begin{table}[tb]
    \centering
    \begin{tabular}{l|l}
    \textbf{Description} & \textbf{Type} \\ 
    \hline
    Boolean truth values & $o$\\
    Statements & $'s$\\
    Sets of statements (e.g. $S$) & $'s\ \texttt{Set}\ :=\ 's\ \ar\ o$\\
    Relations on statements (e.g. $L$) & $'s\ \texttt{Rel}\ :=\ 's\ \ar\ 's\ \ar\ o$\\
    Sets of acceptance conditions (e.g. $C$) & $'s\ \texttt{Cond}\ :=\ 's\ \ar\ 's\ \texttt{Set}\ \ar\ o$\\
    \end{tabular}
\caption{Fundamental base types and type synonyms.}
    \label{tab:typeSynonyms}
\end{table}

A remark is in order. If the type variable of statements $'s$ is instantiated by some concrete type $\alpha$, any term of that type is interpreted by some object from the associated domain $D_{\alpha}$ of the model. However, it need not be the case that the set of statements $S$ contains all objects from $D_\alpha$, but only a subset of those. It is therefore crucial that the encoding is constructed in such a way that any operation or property involving objects of type $'s$ is restricted to the set of statements $S$. This will be addressed throughout the construction of the encoding below.

Basic set-theoretic operations are introduced first. These operations can be defined as anonymous functions built from the underlying HOL logical connectives defined in Sect.~\ref{sec:HOL}. 

\begin{definition}[Basic Set-theoretic Operations]
Set inclusion and set equality are both encoded as functions of type $'s\ \texttt{Set}\ \ar\ 's\ \texttt{Set}\ \ar\ o$ :
\begin{align*}
    &\subseteq\ :=\ \lambda A. \lambda B.\ \forall x.\ (A\ x) \implies (B\ x)\\
    &\approx\ :=\ \lambda A. \lambda B.\ \forall x.\ (A\ x) \iff (B\ x)
\end{align*}
Set intersection is encoded as a function of type $'s\ \texttt{Set}\ \ar\ 's\ \texttt{Set}\ \ar\ 's\ \texttt{Set}$, which corresponds to the type $'s\ \texttt{Set}\ \ar\ 's\ \texttt{Set}\ \ar\ 's\ \ar\ o$ :
\begin{equation*}
    \cap\ :=\ \lambda A. \lambda B. \lambda x.\ (A\ x) \land (B\ x)
\end{equation*}
\end{definition}

In order to increase readability, the above functions are used as infix operators throughout the rest of the article, e.g. $(A \subseteq B)$ formally represents the term $(\subseteq A\ B)$.

The other fundamental concepts that need to be handled by the HOL encoding are two- and three-valued interpretations. 
A two-valued interpretation is a total function $w : S \longrightarrow \{\textbf{t}, \textbf{f}\}$. It therefore seems natural to encode two-valued interpretations as terms $w$ of type $'s\ar o$. However, this does not incorporate the restriction to elements of $S$, as remarked above, into consideration.
In \cite{AFinHOL} so-called "relativised" operators and quantifiers are introduced to address this. 
In this work, however, a restriction on the behaviour of the terms $w$ themselves is imposed. More precisely, a predicate $V_2$ is defined that checks if a term $w$ of type $'s \ar o$ is a valid two-valued interpretation by demanding that all objects of type $'s$ outside of $S$ be mapped to \textbf{f}. This design decision is mainly motivated by two reasons: First and foremost, having a fixed behaviour for objects of type $'s$ outside of $S$ allows for better control of the encoding's behaviour. Secondly, it allows to avoid having to define alternate restricted versions of operations and quantifiers as done in~\cite{AFinHOL}.

The choice of mapping objects of type $'s$ outside of $S$ to \textbf{f} is not arbitrary. Given that sets are encoded through their characteristic function, a two-valued interpretation $w$ (a term of type $'s \ar o$) can thus be seen as a set required to satisfy the property $w \subseteq S$.

\begin{definition}[Two-valued Interpretation]
Given a set of statements $S$, a two-valued interpretation is a term $w$ of type $'s \ar o$ that satisfies the property $w \subseteq S$. This is encoded by the following predicate of type $'s\ \texttt{Set}\ \ar\ ('s \ar o)\ \ar\ o$:
\begin{equation*}
    V_2\ :=\ \lambda S. \lambda w.\ w\subseteq S
\end{equation*}
For convenience a notation for $V_2$ with $S$ already applied is introduced:
\begin{equation*}
    V_2^S\ :=\ V_2(S)
\end{equation*}
$V_2^S$ is a term of type $('s \ar o) \ar o$, and encodes the set of two-valued interpretations over $S$.
\end{definition}

A three-valued interpretation is a total function $v : S \longrightarrow \{\textbf{t},\textbf{f}, \textbf{u}\}$. One possibility to encode $v$ is to assume a new base type $\mathbf{3}$ with domain $D_{\mathbf{3}} = \{\textbf{t},\textbf{f}, \textbf{u}\}$, so that $v$ could be encoded as a term of type $'s \ar \mathbf{3}$. However, this would make it necessary to suitable connect two-valued and three-valued interpretations.
As an alternative, the following representation allows for a natural embedding of two-valued interpretations into the set of three-valued interpretations. More precisely, three-valued interpretations are defined as total functions 
\begin{center}
    $v : X \subseteq S \longrightarrow \{\textbf{t},\textbf{f}\}$,
\end{center}
with elements of $S \setminus X$ being interpreted as \textbf{u}. Three-valued interpretations are thus encoded by two objects, a set $X$ and a mapping $v$. $X$ is then a term of type $'s\ \texttt{Set}$ corresponding to the set containing the statements mapped to either \textbf{t} or \textbf{f} by the interpretation, and $v$ is then a term of type $'s \ar o$ corresponding to the actual mapping of the statements in $X$ to \textbf{t} or \textbf{f}.

The restriction is imposed that every object of type $'s$ outside of $S$ (and $X$) be mapped to \textbf{f}.
A three-valued interpretation is then given by a pair of functions (sets), $X$ and $v$, which must satisfy the property $v \subseteq X \subseteq S$:

\begin{definition}[Three-valued Interpretation]
Given a set of statements $S$, a three-valued interpretation is defined by two terms, $X$ of type $'s\ \texttt{Set}$ and $v$ of type $'s \ar o$, such that they satisfy the property $v \subseteq X \subseteq S$. This is encoded by the following predicate of type $'s\ \texttt{Set}\ \ar\ 's\ \texttt{Set}\ \ar\ ('s \ar o)\ \ar\ o$:
\begin{equation*}
    V_3\ :=\ \lambda S. \lambda X. \lambda v.\ (X \subseteq S) \land (v \subseteq X)
\end{equation*}
Again, a notation for $V_3$ with $S$ already applied is introduced:
\begin{equation*}
    V_3^S\ :=\ V_3(S)
\end{equation*}
$V_3^S$ is a term of type $'s\ \texttt{Set}\ \ar\ ('s \ar o)\ \ar\ o$, and encodes the set of three-valued interpretations over $S$.
\end{definition}

\begin{figure}[tb]
    \centering
    \includegraphics[scale=0.6]{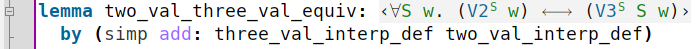}
    \caption{The proof of Lemma \ref{lemma:v2_equiv_v3} in Isabelle/HOL.}
    \label{fig:Isabelle/HOL_proof}
\end{figure}

With the above definitions, the embedding of two-valued interpretations into three-valued interpretations is straight-forward:

\begin{lemma}[\texttt{two\_val\_three\_val\_equiv}]\label{lemma:v2_equiv_v3}
A two-valued interpretation $w$ is a three-valued interpretation $(X, v)$ with $X = S$ and $v = w$:
\begin{equation*}
    \forall S. \forall w.\ (V_2^S\ w) \iff (V_3^S\ S\ w)
\end{equation*} \qed
\end{lemma}
A formal proof within Isabelle/HOL is exemplarily displayed in Fig.~\ref{fig:Isabelle/HOL_proof}.

A final key key concept needs to be encoded: the information ordering $\leq_i$. Recall, for any two three-valued interpretations $v_1$ and $v_2$,
\begin{center}
    $v_1 \leq_i v_2$ \ iff \ $\forall s \in S: v_1(s) \in \{\textbf{t}, \textbf{f}\} \implies v_2(s) = v_1(s)$.
\end{center}
In other words, $v_1 \leq_i v_2$ iff $v_2$ is the same mapping as $v_1$ except with potentially less statements mapped to \textbf{u}. Given the presented encoding of three-valued interpretations, $\leq_i$ can be defined as follows:

\begin{definition}[Information Ordering]
The information ordering $\leq_i$ is encoded as a function of type $'s\ \texttt{Set}\ \ar\ ('s\ar o)\ \ar\ 's\ \texttt{Set}\ \ar\ ('s \ar o)\ \ar\ o$:
\begin{equation*}
    \leq_i\ :=\ \lambda X_1. \lambda v_1. \lambda X_2. \lambda v_2.\ (X_1 \subseteq X_2) \land \Big(\forall s.\ (X_1\ s) \implies \big((v_2\ s) \iff (v_1\ s)\big)\Big)
\end{equation*}
\end{definition}
As usual, $\leq_i$ is used as infix operator throughout the article.

\subsection{Encoding of ADFs Concepts}
Now the encodings of the definitions from Section \ref{sec:ADFs} are introduced. Each definition is first given informally (often leaving the encoded ADF $(S,L,C)$ implicit), and is then followed by its formalisation in HOL.

\begin{definition}[Parent]
Given a set of statements $S$ and a set of links $L$, a term $b$ of type $'s$ is a parent of a term $a$ of type $'s$ iff $(b,a) \in L$ and $b \in S$.
\end{definition}
\begin{formalisation}[Parent] \mbox{} \\
$\texttt{parent}\ :=\ \lambda S. \lambda L. \lambda a. \lambda b.\ (L\ b\ a)\land (S\ b)$\\
where $\texttt{parent}$ is of type $'s\ \texttt{Set}\ \ar\ 's\ \texttt{Rel}\ \ar\ 's\ \ar\ 's\ \ar\ o$.\\
\\
$\texttt{parent}^{S,L}\ :=\ \texttt{parent}(S)(L)$\\
where $\texttt{parent}$ is of type $'s\ \ar\ 's\ \ar\ o$.
\end{formalisation}

In the definition above, 
the constraint $b \in S$ is, again, due to the technical reason that it must be ensured that every operation performed by the encoding is restricted to $S$. 
It should be noted that the definition above implicitly assumes that the input $a$ of type $'s$ be such that $a \in S$. This constraint is implemented through the operator $G$ defined further below.

In order to verify that our definition of $\texttt{parent}$ behaves properly, the following inclusion is proven using Isabelle/HOL:

\begin{lemma}[\texttt{parent\_subset\_S}]
$\forall S. \forall L. \forall s.\ (\texttt{parent}^{S,L}\ s) \subseteq S$ \qed
\end{lemma}

\noindent Intuitively, the set of parents of a statement $s$ is always a subset of the set of statements $S$. The opposite inclusion $\forall S. \forall L. \forall s.\ S \subseteq (\texttt{parent}^{S,L}\ s)$ is automatically disproved by a counter-model using Nitpick~\cite{files}. 

Before defining the two-valued model operator $G$, a last technical issue needs to be resolved. Similarly to the set of links $L$, the set of acceptance conditions $C$ is encoded without any further restrictions other than its type $'s\ \texttt{Cond}$. And in particular, for any statement $s$, $(C\ s)$ is a term of type $'s\ \texttt{Set} \ar o$. This means that the representation admits models with a term $(C\ s)$ that also provides a mapping for sets that are not part of the power set of the parents of $s$; unlike what is required by Definition \ref{def:ADF}. 

To solve this issue, the function $\texttt{restricted\_C}$ is introduced that builds a proper set of acceptance conditions out of a term $C$ of type $'s\ \texttt{Cond}$:

\begin{definition}[Set of Acceptance Conditions]
A term $C$ of type $'s\ \texttt{Cond}$ is a set of acceptance conditions iff, for every statement $s$, $(C\ s)$ is a function mapping to \textbf{t} only subsets of the set of parents of $s$.
\end{definition}
\begin{formalisation}[Set of Acceptance Conditions] \mbox{} \\
$\texttt{restricted\_C}\ :=\ \lambda S. \lambda L. \lambda C. \lambda s. \lambda X.\ \exists A.\ (C\ s\ A) \land \big((A \cap \texttt{parent}^{S,L}\ s) \approx X\big)$\\
of type $'s\ \texttt{Set}\ \ar\ 's\ \texttt{Rel}\ \ar\ 's\ \texttt{Cond}\ \ar\ 's\ \texttt{Cond}$.\\
\\
$\texttt{C}^{S,L,C}\ :=\ \texttt{restricted\_C}(S)(L)(C)$\\
of type $'s\ \texttt{Cond}$.
\end{formalisation}

\noindent Intuitively, for every statement $s$, the function $\texttt{restricted\_C}$ gives sets $A$ mapped to \textbf{t} by $(C\ s)$ and restricts them to the set of parents of $s$, i.e.\ $(A \cap \texttt{parent}^{S,L}\ s)$. The newly built set of acceptance conditions $\texttt{C}^{S,L,C}$ then validates the following property:

\begin{lemma}[\texttt{restricted\_C}] \mbox{} \\
$\forall S. \forall L. \forall C. \forall s. \forall X.\ (\texttt{C}^{S,L,C}\ s\ X) \implies (X \subseteq \texttt{parent}^{S,L}\ s)$ \qed
\end{lemma}
\noindent If a set $X$ is mapped to \textbf{t} by the function $(\texttt{C}^{S,L,C}\ s)$, then $X$ is a subset of parents of $s$. The converse is again automatically disproved by a counter-model.

Now the two-valued model operator $G$ of Def.~\ref{def:G_operator} is encoded:

\begin{definition}[Two-valued Model Operator]
The two-valued model operator $\texttt{G}$ takes as input a two-valued interpretation $w$ and outputs an updated two-valued interpretation that maps each statement $s$ to its evaluated acceptance condition under $w$.
\end{definition}
\begin{formalisation}[Two-valued Model Operator] \mbox{} \\
$\texttt{G}\ :=\ \lambda S. \lambda L. \lambda C. \lambda w. \lambda s.\ (S\ s) \land \big(\texttt{C}^{S,L,C}\ s\ (w \cap \texttt{parent}^{S,L}\ s)\big)$\\
of type $'s\ \texttt{Set}\ \ar\ 's\ \texttt{Rel}\ \ar\ 's\ \texttt{Cond}\ \ar\ ('s \ar o)\ \ar\ ('s \ar o)$.\\
\\
$\texttt{G}^{S,L,C}\ :=\ \texttt{G}(S)(L)(C)$\\
of type $('s \ar o)\ \ar\ ('s \ar o)$.
\end{formalisation}

\noindent In the above definition, the constraint $s \in S$, encoded by $(S \; s)$, is added for two reasons. First, it restricts the use of the functions $\texttt{parent}^{S,L}$ and $\texttt{C}^{S,L,C}$ to objects of type $'s$ that are elements of the set of statements $S$. Secondly, it is a requirement of the encoding for $(\texttt{G}^{S,L,C}\ w)$ indeed returning a two-valued interpretation. Recall, for the term $(\texttt{G}^{S,L,C}\ w)$ of type $('s \ar o)$ to be a two-valued interpretation, it should hold that $(\texttt{G}^{S,L,C}\ w) \subseteq S$. This is indeed the case (with the converse again being disproved by Nitpick):

\begin{lemma}[\texttt{two\_val\_operator}]
$\forall S. \forall L. \forall C. \forall w.\ V_2^S\ (\texttt{G}^{S,L,C}\ w)$\\
Equivalently: $\forall S. \forall L. \forall C. \forall w. \forall s.\ \neg(S\ s) \implies \big((\texttt{G}^{S,L,C}\ w\ s) \iff \bot\big)$ \qed
\end{lemma}

With the two-valued model operator $\texttt{G}^{S,L,C}$ defined, it is now possible to encode two-valued models:

\begin{definition}[Two-valued Model]
A term $w$ of type $('s \ar o)$ is a two-valued model iff $w$ is a two-valued interpretation and a fixpoint of operator $\texttt{G}^{S,L,C}$.
\end{definition}
\begin{formalisation}[Two-valued Model] \mbox{} \\
$\texttt{model}\ :=\ \lambda S. \lambda L. \lambda C. \lambda w.\ (V_2^S\ w) \land \big((\texttt{G}^{S,L,C}\ w) \approx w\big)$\\
of type $'s\ \texttt{Set}\ \ar\ 's\ \texttt{Rel}\ \ar\ 's\ \texttt{Cond}\ \ar\ ('s \ar o)\ \ar\ o$.\\
\\
$\texttt{model}^{S,L,C}\ :=\ \texttt{model}(S)(L)(C)$\\
of type $('s \ar o)\ \ar\ o$.
\end{formalisation}

\noindent By definition, it is straight-forward that any two-valued model is a two-valued interpretation (the converse does not hold):

\begin{lemma}[\texttt{two\_val\_model}] \mbox{} \\
$\forall S. \forall L. \forall C. \forall w.\ (\texttt{model}^{S,L,C}\ w) \implies (V_2^S\ w)$. \qed
\end{lemma}

In order to encode the ultimate approximation operator $\Gamma$ from Def.~\ref{def:Gamma_operator}, the notion of a two-valued interpretation approximated by a three-valued interpretation is encoded first.
Recall, a single three valued interpretation $v : S \longrightarrow \{\textbf{t}, \textbf{f}, \textbf{u}\}$ approximates a set of two-valued interpretations defined by
\begin{center}
    $[v]_2 := \{w \in V_2^S\ |\ v \leq_i w\}$,
\end{center}
where $V_2^S$ denotes the set of two-valued interpretations over $S$. And given the presented encoding of two- and three-valued interpretations, the above notion is encoded as follows:

\begin{definition}[Set of Approximated Two-valued Interpretations]
The set of two-valued interpretations approximated by a three-valued interpretation, denoted $[X,v]_2^S$, is composed of the two-valued interpretations with more information content than the approximating three-valued interpretation $(X, v)$.
\end{definition}
\begin{formalisation}[Set of Approximated Two-valued Interpretations] \mbox{} \\
$\texttt{approx\_two\_val}\ :=\ \lambda X. \lambda v. \lambda S. \lambda w.\ (V_2^S\ w) \land (X\ v \leq_i S\ w)$\\
of type $'s\ \texttt{Set}\ \ar\ ('s \ar o)\ \ar\ 's\ \texttt{Set}\ \ar\ ('s \ar o)\ \ar\ o$.\\
\\
$[X,v]_2^S\ :=\ \texttt{approx\_two\_val}(X)(v)(S)$\\
of type $('s \ar o)\ \ar\ o$.
\end{formalisation}

\noindent By definition it is evident that any element of $[X,v]_2^S$ is a two-valued interpretation, as verified by Isabelle/HOL:

\begin{lemma}[\texttt{approx\_two\_val\_V2}] \mbox{} \\
$\forall S. \forall X. \forall v. \forall w.\ ([X,v]_2^S\ w) \implies (V_2^S\ w).$ \qed
\end{lemma}

\noindent Finally, it should be noted that if the approximating three-valued interpretation $(X, v)$ is itself a two-valued interpretation $w$ (i.e.\ $X = S$ and $v = w$), then the set of approximated two-valued interpretations $[X,v]_2^S$ contains a  single element, i.e. $v$ itself.
In fact, this property is equivalent to the definition of two-valued interpretations (here $(\lambda w'.\ w' \approx w)$ encodes the set of two-valued interpretations containing only $w$):

\begin{lemma}[\texttt{V2\_approx\_two\_val\_equiv}] \mbox{} \\
$\forall S. \forall w.\ (V_2^S\ w)\iff \big([S,w]_2^S \approx (\lambda w'.\ w' \approx w)\big)$ \qed
\end{lemma}

The ultimate approximation operator $\Gamma$ takes as input a three-valued interpretation and outputs a revised three-valued interpretation. Since three-valued interpretations are encoded by pairs $(X,v)$, the operator $\Gamma$ also needs to be split into two terms: a term $\Gamma_X$ and a term $\Gamma_v$. The pair $\big(\Gamma_X(X,v),\ \Gamma_v(X,v)\big)$ then defines the updated three-valued interpretation obtained by applying the operator $\Gamma$ on $(X,v)$. More precisely, $\Gamma_X(X,v)$ corresponds to the set containing the statements mapped to either \textbf{t} or \textbf{f} by the updated interpretation, and $\Gamma_v(X,v)$ corresponds to the actual mapping of the statement in $\Gamma_X(X,v)$ to \textbf{t} or \textbf{f}. Furthermore, the set $S \setminus \Gamma_X(X,v)$ then corresponds to the set of statements mapped to \textbf{u} by the updated interpretation.

\begin{definition}[Ultimate Approximation Operator]
The ultimate approximation operator $\Gamma$ takes as input a three-valued interpretation $(X, v)$ and outputs an updated three-valued interpretation that maps each statement $s$ to the value of the consensus among all two-valued interpretations $w \in [X,v]_2^S$ concerning the evaluation of the statement's acceptance condition under each $w$.
\end{definition}
\begin{formalisation}[Ultimate Approximation Operator] 
\begin{flalign*}
\Gamma_X := &\lambda S. \lambda L.\lambda C. \lambda X. \lambda v. \lambda s.\\ &(S\ s) \land \Big(\forall w1. \forall w2.\ \big(([X,v]_2^S\ w1) \land ([X,v]_2^S\ w2)\big) \implies && \\ & \hspace*{3cm} \big((\texttt{G}^{S,L,C}\ w1\ s) \iff (\texttt{G}^{S,L,C}\ w2\ s)\big)\Big)
\end{flalign*}
of type $'s\ \texttt{Set}\ \ar\ 's\ \texttt{Rel}\ \ar\ 's\ \texttt{Cond}\ \ar\ 's\ \texttt{Set}\ \ar\ ('s \ar o)\ \ar\ 's\ \texttt{Set}$.\\
\\
$\Gamma_X^{S,L,C} := \Gamma_X(S)(L)(C)$\\
of type $'s\ \texttt{Set}\ \ar\ ('s \ar o)\ \ar\ 's\ \texttt{Set}$.
\begin{flalign*}
\Gamma_v := \lambda S. \lambda L.&\lambda C. \lambda X. \lambda v. \lambda s.\, \\ & (S \; s) \land \Big(\forall w.\ ([X,v]_2^S\ w) \implies \big((\texttt{G}^{S,L,C}\ w\ s) \iff \top \big)\Big) &
\end{flalign*}
of type $'s\ \texttt{Set}\ \ar\ 's\ \texttt{Rel}\ \ar\ 's\ \texttt{Cond}\ \ar\ 's\ \texttt{Set}\ \ar\ ('s \ar o)\ \ar\ ('s \ar o)$.\\
\\
$\Gamma_v^{S,L,C} := \Gamma_v(S)(L)(C)$\\
of type $'s\ \texttt{Set}\ \ar\ ('s \ar o)\ \ar\ ('s \ar o)$.
\end{formalisation}

\noindent In the definition above, the constraint $s \in S$ is added for the exact same reasons as for operator $\texttt{G}$, as well as for consistency with the previous encoding of three-valued interpretations, i.e. that $(\Gamma_v^{S,L,C}\ X\ v) \subseteq (\Gamma_X^{S,L,C}\ X\ v) \subseteq S$. Additionally, given the definition of operator $\texttt{G}$, the term $(\texttt{G}^{S,L,C}\ w\ s)$ corresponds to the evaluation of acceptance condition $(\texttt{C}^{S,L,C}\ s)$ of statement $s$ under interpretation $w$.

The encoding of operator $\Gamma$ guarantees that the pair 
$$\big((\Gamma_X^{S,L,C}\ X\ v),\ (\Gamma_v^{S,L,C}\ X\ v)\big)$$
 describes a three-valued interpretation, formally:

\begin{lemma}[\texttt{ult\_approx\_operator\_V3}] \mbox{} \\
$\forall S. \forall L. \forall C. \forall X. \forall v.\ V_3^S\ (\Gamma_X^{S,L,C}\ X\ v)\ (\Gamma_v^{S,L,C}\ X\ v)$ \qed
\end{lemma}

In order to further verify that the presented encoding of operator $\Gamma$ indeed behaves as expected, the situation is examined where a statement $s$ is mapped to \textbf{f} by the updated three-valued interpretation $\big((\Gamma_X^{S,L,C}\ X\ v),\ (\Gamma_v^{S,L,C}\ X\ v)\big)$. In that case, it should be the case that $s \in (\Gamma_X^{S,L,C}\ X\ v)$ but $s \notin (\Gamma_v^{S,L,C}\ X\ v)$. This should also correspond to the situation where the consensus truth value of the evaluation of acceptance condition $(\texttt{C}^{S,L,C}\ s)$ under each $w \in [X,v]_2^S$ is \textbf{f}. This is verified by Isabelle/HOL:

\begin{lemma}[\texttt{ult\_approx\_operator\_bot\_consensus}] 
\begin{flalign*}
\forall S. \forall L. \forall C. \forall X. &\forall v. \forall s.\ (V_3^S\ X\ v) \implies \\ &\Bigg( \Big((\Gamma_X^{S,L,C}\ X\ v\ s) \land \neg(\Gamma_v^{S,L,C}\ X\ v\ s)\Big) \iff \\ & \quad\bigg((S\ s) \land \Big(\forall w.\ ([X,v]_2^S\ w) \implies \big((\texttt{G}^{S,L,C}\ w\ s) \iff \bot \big)\Big)\bigg) \ \Bigg)
\end{flalign*} \qed
\end{lemma}

Another important property is the behaviour of operator $\Gamma$ when given a two-valued interpretation as input. Recall that if $w$ is a two-valued interpretation then $[w]_2 = \{w\}$, and thus $\Gamma_D(w)(s) = w(\varphi_s) = G_D(w)(s)$. This is indeed the case, as verified by Isabelle/HOL (again, the converses of the two properties were automatically disproved by Nitpick):

\begin{lemma} \mbox{}
\begin{enumerate} \itemsep.5em
  \item  (\texttt{V2\_$\Gamma$X\_G\_relation}).\\
$\forall S. \forall L. \forall C. \forall w.\ (V_2^S\ w) \implies \big((\Gamma_X^{S,L,C}\ S\ w) \approx S\big)$
  \item (\texttt{V2\_$\Gamma$v\_G\_relation}).\\
$\forall S. \forall L. \forall C. \forall w.\ (V_2^S\ w) \implies \big((\Gamma_v^{S,L,C}\ S\ w) \approx (\texttt{G}^{S,L,C}\ w)\big)$
\end{enumerate}
\qed
\end{lemma}

\noindent Finally, an important and well-know property of operator $\Gamma$ is its monotonicity with respect to the information ordering $\leq_i$. This is again formally proven using Isabelle/HOL:

\begin{lemma}[\texttt{$\Gamma$\_info\_ordering\_monotone}]\label{lemma:Gamma_monotonicity} \mbox{} \\
$\forall S. \forall L. \forall C. \forall X. \forall v. \forall X'. \forall v'.\ \Big((V_3^S\ X\ v) \land (V_3^S\ X'\ v') \land (X\ v \leq_i X'\ v')\Big) \implies \\ \Big((\Gamma_X^{S,L,C}\ X\ v)\ (\Gamma_v^{S,L,C}\ X\ v) \leq_i (\Gamma_X^{S,L,C}\ X'\ v')\ (\Gamma_v^{S,L,C}\ X'\ v')\Big)$ \qed
\end{lemma}

\noindent Now the standard ADF semantics are encoded.

\begin{definition}[Admissible, Complete, Preferred and Grounded Semantics]\label{def:standard_semantics_encoding}
Let (X,v) be a three-valued interpretation.
\begin{enumerate}
    \item $(X,v)$ is \textit{admissible} iff $(X,v) \leq_i \Gamma(X,v)$.
    \item $(X,v)$ is \textit{complete} iff $(X,v) = \Gamma(X,v)$.
    \item $(X,v)$ is \textit{preferred} iff $(X,v)$ is $\leq_i$-maximal admissible.
    \item $(X,v)$ is \textit{grounded} iff $(X,v)$ is the $\leq_i$-least fixpoint of $\Gamma$.
\end{enumerate}
\end{definition}
\begin{formalisation}[Adm., Complete, Pref.\ and Grounded Semantics] \mbox{} 
\begin{enumerate}\itemsep.5em
  \item $\begin{aligned}[t]\texttt{admissible} := \lambda S. \lambda L.\lambda C. &\lambda X. \lambda v.  (V_3^S \; X \; v) \; \land \\
  &\Big(X \; v \leq_i (\Gamma_X^{S,L,C} \, X \; v) \; (\Gamma_v^{S,L,C} \; X \; v)\Big)\end{aligned}$\\
of type: $'s\ \texttt{Set}\ \ar\ 's\ \texttt{Rel}\ \ar\ 's\ \texttt{Cond}\ \ar\ 's\ \texttt{Set}\ \ar\ ('s \ar o)\ \ar\ o$.\\

$\texttt{admissible}^{S,L,C} := \texttt{admissible}(S)(L)(C)$\\
of type: $'s\ \texttt{Set}\ \ar\ ('s \ar o)\ \ar\ o$.
  \item $\begin{aligned}[t]\texttt{complete} := \lambda S. \lambda L.\lambda C. &\lambda X. \lambda v.\  (V_3^S\ X\ v) \; \land \\
  &\Big((\Gamma_X^{S,L,C}\ X\ v) \approx X\Big) \land \Big((\Gamma_v^{S,L,C}\ X\ v) \approx v \Big)\end{aligned}$\\
of type: $'s\ \texttt{Set}\ \ar\ 's\ \texttt{Rel}\ \ar\ 's\ \texttt{Cond}\ \ar\ 's\ \texttt{Set}\ \ar\ ('s \ar o)\ \ar\ o$.\\

$\texttt{complete}^{S,L,C}\ :=\ \texttt{complete}(S)(L)(C)$\\
of type: $'s\ \texttt{Set}\ \ar\ ('s \ar o)\ \ar\ o$.
  \item $\begin{aligned}[t]\texttt{preferred} := &\lambda S. \lambda L.\lambda C. \lambda X. \lambda v. (\texttt{admissible}^{S,L,C}\ X\ v) \; \land \\ 
  &\bigg(\forall X'. \forall v'.\ \begin{aligned}[t]&\Big((\texttt{admissible}^{S,L,C}\ X'\ v') \land (X\ v \leq_i X'\ v')\Big) \implies \\
  &\Big((X \approx X') \land (v \approx v')\Big) \bigg)\end{aligned}\end{aligned}$\\
of type: $'s\ \texttt{Set}\ \ar\ 's\ \texttt{Rel}\ \ar\ 's\ \texttt{Cond}\ \ar\ 's\ \texttt{Set}\ \ar\ ('s \ar o)\ \ar\ o$.\\

$\texttt{preferred}^{S,L,C}\ :=\ \texttt{preferred}(S)(L)(C)$\\
of type: $'s\ \texttt{Set}\ \ar\ ('s \ar o)\ \ar\ o$.
  \item $\begin{aligned}[t]\texttt{grounded}\ :=\ &\lambda S. \lambda L.\lambda C. \lambda X. \lambda v.\ (\texttt{complete}^{S,L,C}\ X\ v) \; \land \\ & \Big(\forall X'. \forall v'.\ (\texttt{complete}^{S,L,C}\ X'\ v') \implies (X\ v \leq_i X'\ v')\Big)\end{aligned}$\\
of type: $'s\ \texttt{Set}\ \ar\ 's\ \texttt{Rel}\ \ar\ 's\ \texttt{Cond}\ \ar\ 's\ \texttt{Set}\ \ar\ ('s \ar o)\ \ar\ o$.\\

$\texttt{grounded}^{S,L,C}\ :=\ \texttt{grounded}(S)(L)(C)$\\
of type: $'s\ \texttt{Set}\ \ar\ ('s \ar o)\ \ar\ o$.
\end{enumerate}
\end{formalisation}

\noindent Before encoding the stable semantics, the notion of reduct of an ADF $(S,L,C)$, see Def.~\ref{def:stable_semantics}, needs to be encoded.

\begin{definition}[Reduct of $L$ and $C$]
Let $w$ be a two-valued interpretation.
\begin{enumerate}
    \item The reduct of the set of links $L$ is obtained by reducing $L$ to the set of links between statements mapped to \textbf{t} by $w$.
    \item The reduct of the set of acceptance conditions $C$ is obtained by updating each acceptance condition $(C\ s)$ in such a way that the subsets $X$ of parents of $s$ used to describe the mapping $(C\ s)$ be limited only to those where the statements mapped to \textbf{f} by $w$ are not included. Furthermore, given that the reduct of the set of statements $S$ is by definition the set of statements mapped to \textbf{t} by $w$, the above modification for $C$ only matters for those statements mapped to \textbf{t} by $w$.
\end{enumerate}
\end{definition}
\begin{formalisation}[Reduct of $L$ and $C$] \mbox{} 
\begin{enumerate}\itemsep.5em
  \item $\texttt{reduct\_L}\ :=\ \lambda L. \lambda w. \lambda s1. \lambda s2.\ (L\ s1\ s2) \land (w\ s1) \land (w\ s2)$\\
Type: $'s\ \texttt{Rel}\ \ar\ ('s \ar o)\ \ar\ 's\ \texttt{Rel}$
  \item $\texttt{reduct\_C}\ :=\ \lambda C. \lambda w. \lambda s. \lambda X.\ (w\ s) \land (C\ s\ X) \land \big(\forall s'.\ \neg (w\ s') \implies \neg (X\ s')\big)$\\
Type: $'s\ \texttt{Cond}\ \ar\ ('s \ar o)\ \ar\ 's\ \texttt{Cond}$
\end{enumerate}
\end{formalisation}

\noindent The encoding of stable model semantics is presented next.

\begin{definition}[Stable Model Semantics]
A two-valued model $w$ of an ADF $D$ is stable iff the reduct of $D$ is an ADF for which the grounded interpretation maps to \textbf{t} all statements of the reduct (i.e.\ statements  mapped to True by $w$).
\end{definition}
\begin{formalisation}[Stable Model Semantics] \mbox{} \\
$\begin{aligned}[t]
\texttt{stable}\ :=\ \lambda S. \lambda L.\lambda C. \lambda w.\ &(\texttt{model}^{S,L,C}\ w) \; \land \\ & \big(\texttt{grounded}\ w\ (\texttt{reduct\_L}\ L\ w)\ (\texttt{reduct\_C}\ C\ w)\ w\ w\big)
\end{aligned}$\\
of type: $'s\ \texttt{Set}\ \ar\ 's\ \texttt{Rel}\ \ar\ 's\ \texttt{Cond}\ \ar\ ('s\ar o)\ \ar\ o$.\\
\\
$\texttt{stable}^{S,L,C}\ :=\ \texttt{stable}(S)(L)(C)$\\
of type: $('s\ar o)\ \ar\ o$.
\end{formalisation}

\noindent In the above definition, the first three inputs of the function \texttt{grounded} correspond to the reduct of the ADF $(S,L,C)$ with respect to the two-valued model $w$. The last two inputs of the function \texttt{grounded} correspond to the three-valued interpretation of the reduct where all statements are mapped to \textbf{t}.

The last step of the encoding concerns the special case where an ADF $(S,L,C)$ describes a simple AF. Recall, an AF $(A,R)$ can be described by an ADF $(A,R,C)$ such that every acceptance condition $(C\ s)$ only maps the empty set to \textbf{t}.
Given this definition, a predicate is encoded that checks if the set of acceptance conditions $C$ of an ADF $(S,L,C)$ describes an AF:

\begin{definition}[Acceptance Conditions of an AF]\label{def:acceptance_condition_AFs}
The set of acceptance conditions $C$ of an ADF $(S,L,C)$ describes an AF iff every acceptance condition (C\ s) only maps the empty set to \textbf{t}.
\end{definition}
\begin{formalisation} \mbox{} \\
$\texttt{C\_AF}\ :=\ \lambda C.\ \forall s. \forall X.\ (C\ s\ X) \iff \big(X \approx (\lambda x.\ \bot)\big)$, \\
of type: $'s\ \texttt{Cond}\ \ar\ o$, where $(\lambda x.\ \bot)$ corresponds to the empty set.
\end{formalisation}

\section{Meta-Theoretical Properties and Application Examples}

In this section the encoding of ADFs in HOL is employed for (a) verifying meta-theoretical properties, and (b) for conducting an analysis of two concrete argumentation frameworks. Again, every valid property is proven using Isabelle/HOL (stating the name of the lemma in the theory source files), and counter-examples to statements that should not hold are automatically generated by Nitpick.

\subsection{Meta-Theory}\label{sec:semantic_properties}
A first important property is the uniqueness of grounded interpretations. Even though the definition of \texttt{grounded}, see Def.\ \ref{def:standard_semantics_encoding}, should guarantee the uniqueness of grounded interpretations, it remained to be verified. It is given by:

\begin{lemma}[\texttt{grounded\_superscript\_uniqueness}] \mbox{} \\
$\forall S. \forall L. \forall C. \forall X. \forall v.\ (\texttt{grounded}^{S,L,C}\ X\ v) \implies \\ \bigg(\forall X'. \forall v'.\ \Big((V_3^S\ X'\ v') \land \big(\neg (X' \approx X) \lor \neg (v' \approx v)\big)\Big) \implies \neg (\texttt{grounded}^{S,L,C}\ X'\ v')\bigg)$ \qed
\end{lemma}

\begin{figure}[tb]
    \centering
    \includegraphics[width=0.7\textwidth]{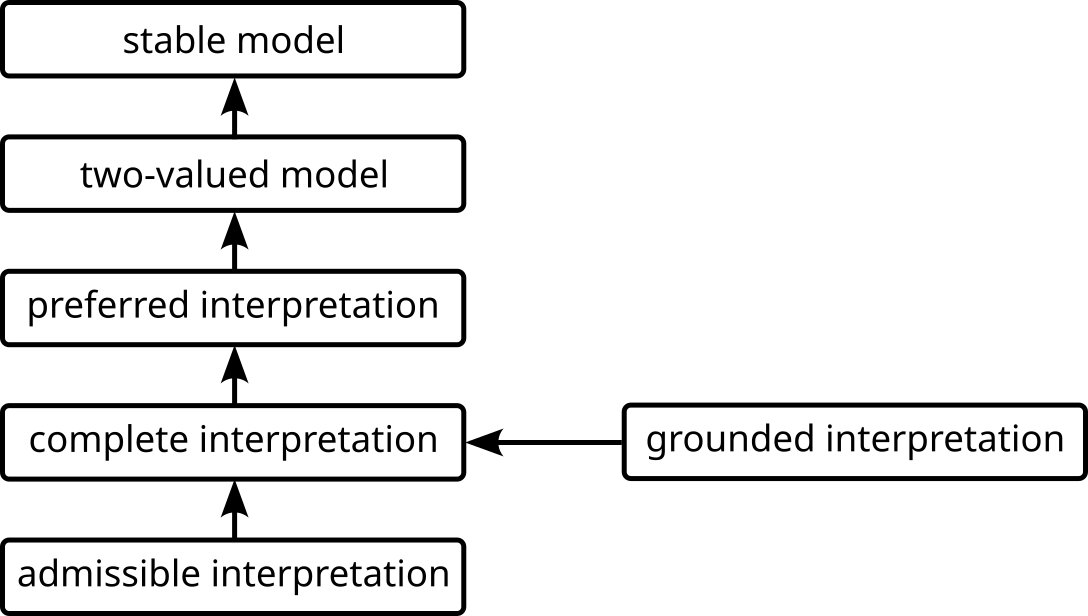}
    \caption{Relations between ADF semantics, following~\cite{ADFs}. Arrows represent inclusion relations, transitive arrows are omitted.}
    \label{fig:ADF_semantics_relations_ENCODING}
\end{figure}

One of the most fundamental results the encoding should provide is the establishment of the various relationships between standard ADF semantics, abstractly stated in Theorem \ref{theorem:relation_between_ADF_semantics} and summarised in Fig.~\ref{fig:ADF_semantics_relations_ENCODING}.
This series of implications is encoded as follows, each of which being formally verified by Isabelle/HOL.

\begin{theorem}\label{theorem:relation_between_ADF_semantics_ENCODING} \mbox{} 
\begin{enumerate} \itemsep0.5em
\item (\texttt{stable\_to\_model}).\\
$\forall S. \forall L. \forall C. \forall w.\ \texttt{stable}^{S,L,C}\ w \implies \texttt{model}^{S,L,C}\ w$
\item (\texttt{model\_to\_preferred}).\\
$\forall S. \forall L. \forall C. \forall w.\ \texttt{model}^{S,L,C}\ w \implies \texttt{preferred}^{S,L,C}\ S\ w$
\item (\texttt{preferred\_to\_complete}).\\
$\forall S. \forall L. \forall C. \forall X. \forall v.\ \texttt{preferred}^{S,L,C}\ X\ v \implies \texttt{complete}^{S,L,C}\ X\ v$
\item (\texttt{complete\_to\_admissible}).\\
$\forall S. \forall L. \forall C. \forall X. \forall v.\ \texttt{complete}^{S,L,C}\ X\ v \implies \texttt{admissible}^{S,L,C}\ X\ v$
\item (\texttt{grounded\_to\_complete}).\\
$\forall S. \forall L. \forall C. \forall X. \forall v.\ \texttt{grounded}^{S,L,C}\ X\ v \implies \texttt{complete}^{S,L,C}\ X\ v$
\end{enumerate}\qed
\end{theorem}

\begin{figure}[tb]
    \centering
    \begin{subfigure}{.65\textwidth}
    \centering
    \includegraphics[width=\textwidth]{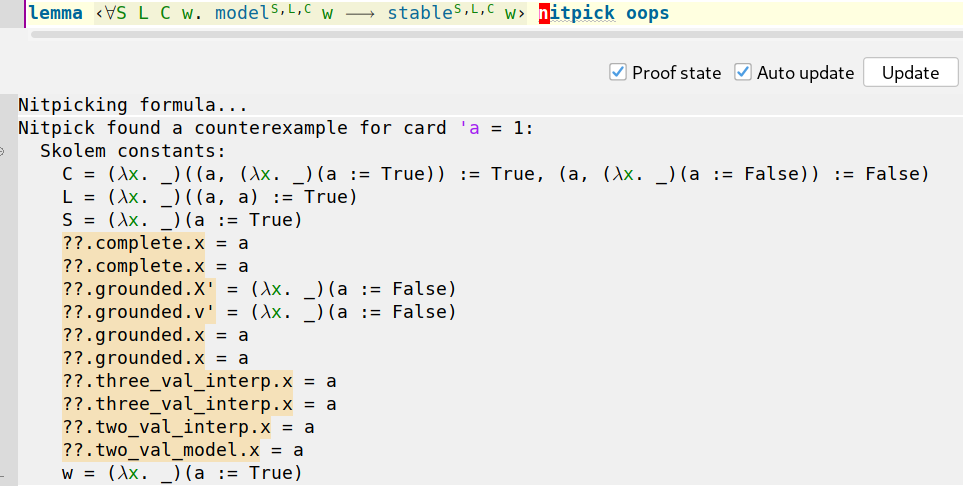}
    \caption{System output\label{fig:nitpick:ex1}}
    \end{subfigure}\;
    \begin{subfigure}{.3\textwidth}
    \centering
    \includegraphics[width=.35\textwidth]{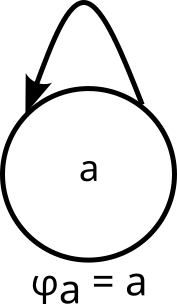}
    \caption{Visualization\label{fig:nitpick:ex2}}
    \end{subfigure}
    
    \caption{Counter-example generation using Nitpick for an invalid inclusion statement.}
    \label{fig:nitpick-counterexamples-relation}
\end{figure}

Additionally, the converses of all the above properties were disproved via counter-models automatically generated by Nitpick.
An example for such an output, here for the non-inclusion of two-valued models in the set of stable models (opposite direction of Theorem~\ref{theorem:relation_between_ADF_semantics_ENCODING}~(1.)), is displayed in Fig.~\ref{fig:nitpick:ex1}. The output of Nitpick
is rather technical, a visualization of the output as a simple ADF is given in Fig.~\ref{fig:nitpick:ex2}. Intuitively, each line of
of the Nitpick output defines a higher-order constant symbol (such as $C$, $L$ and $S$) that describes the concrete ADF used in the
counter-example. For example, the expression \lstinline[mathescape,basicstyle=\ttfamily]|L = ($\lambda$x. _)((a,a) := True)| defines
the encoded set of links $L$ to be a set of one element $\{(a,a) \}$ (represented by its characteristic function). The
two-valued model that is not a stable model is given by the witness \lstinline[mathescape,basicstyle=\ttfamily]|w = ($\lambda$x. _)(a := True)|,
i.e., the two-valued model that maps argument $a$ to \textbf{t}.
An in-depth explanation of Nitpick's syntax
is provided in its documentatation\footnote{\url{https://isabelle.in.tum.de/dist/Isabelle2023/doc/nitpick.pdf}}.

In the above theorem, the assertion \texttt{preferred\_to\_complete} proved to be the most challenging and required an intermediate result in order to be formally verified. This intermediate result is a well-known lemma which can be derived from the monotonicity of the $\Gamma$ operator (with respect to the information ordering $\leq_i$), cf.\ Lemma \ref{lemma:Gamma_monotonicity}. It states that the updated three-valued interpretation $\Gamma(X,v)$ of an admissible three-valued interpretation $(X,v)$ is also admissible:

\begin{lemma}[\texttt{preferred\_to\_complete\_1}]
\begin{flalign*}
\forall S. \forall L. \forall C. \forall X. \forall v.\ & \texttt{admissible}^{S,L,C}\ X\ v \implies & \\
  &\texttt{admissible}^{S,L,C}\ (\Gamma_X^{S,L,C}\ X\ v)\ (\Gamma_v^{S,L,C}\ X\ v) & 
\end{flalign*} \qed
\end{lemma}

Finally Theorem \ref{theorem:relation_between_ADF_semantics_ENCODING} can be extended to show that, in the case of two-valued interpretations, the concepts of \texttt{model}, \texttt{preferred} and \texttt{complete} are equivalent:

\begin{lemma} \mbox{}
\begin{enumerate}\itemsep.5em
\item (\texttt{V2\_model\_equiv\_preferred}).\\
$\forall S. \forall L. \forall C. \forall w.\ \texttt{model}^{S,L,C}\ w \iff \texttt{preferred}^{S,L,C}\ S\ w$
\item (\texttt{V2\_model\_equiv\_complete}).\\
$\forall S. \forall L. \forall C. \forall w.\ \texttt{model}^{S,L,C}\ w \iff \texttt{complete}^{S,L,C}\ S\ w$
\end{enumerate} \qed
\end{lemma}

\noindent Furthermore, it holds that there always exists at least one admissible interpretation given any ADF $(S,L,C)$:
\begin{equation*}
    \forall S. \forall L. \forall C.\ \exists X. \exists v.\ \texttt{admissible}^{S,L,C}\ X\ v 
\end{equation*}

Additionally, Nitpick is able to generate a counter-model for the following property:
\begin{equation*}
    \forall S. \forall L. \forall C. \forall X. \forall v.\ \texttt{admissible}^{S,L,C}\ X\ v
\end{equation*}

Intuitively, the formula states that, given an ADF $(S,L,C)$, not every three-valued interpretation $(X,v)$ is admissible. And given the chain of implications from Theorem \ref{theorem:relation_between_ADF_semantics_ENCODING} this result also applies to all other semantics defined here.

\medskip

The last important result discussed in this article concerns the equivalence between the concepts of \texttt{model} and \texttt{stable} for argumentation frameworks (AFs). Recall that an AF $(A,R)$ can be encoded as an ADF $(S,L,C)$ by setting $S = A$, $L = R$, and by demanding that $C$ follows Def.~\ref{def:acceptance_condition_AFs}, i.e., that $C$ be such that every acceptance condition $(C\ s)$ only maps the empty set to \textbf{t}.

Proving this equivalence within Isabelle/HOL is more challenging and cannot be done automatically by Sledgehammer; some fundamental properties relating to AFs have to be formally verified as lemmas first. For reasons of brevity, these lemmas are omitted here but can be found in the Isabelle/HOL source files. Using these lemmas the following theorem can then be verified.

\begin{theorem}[\texttt{model\_equiv\_stable\_AF}] \mbox{} \\
$\forall S. \forall L. \forall C. \forall w.\ (\texttt{C\_AF}\ C) \implies \Big((\texttt{model}^{S,L,C}\ w) \iff (\texttt{stable}^{S,L,C}\ w)\Big)$
\qed
\end{theorem}

When removing the constraint $(\texttt{C\_AF}\ C)$ which imposes that the ADF $(S,L,C)$ represent an AF, the above property is automatically disproved by Nitpick.

\subsection{Application Examples}
Two application examples are given that illustrate the ability of the presented encoding to not only reason about ADFs (and AFs) but
also to compute or formalize extensions of given ADFs (AFs). First, Example \ref{ex:ADF} of Section \ref{sec:ADFs} is used illustrate the encoding of ADFs in general, and then a (simple) AF example illustrates the application of the same encoding to AFs. The satisfiability of each property mentioned below is either confirmed or disproved by the Nitpick model-finder.

\subsubsection{Example ADF.}
\begin{figure}[tb]
\centering
\begin{minipage}{.45\textwidth}
  \includegraphics[height=5cm,interpolate]{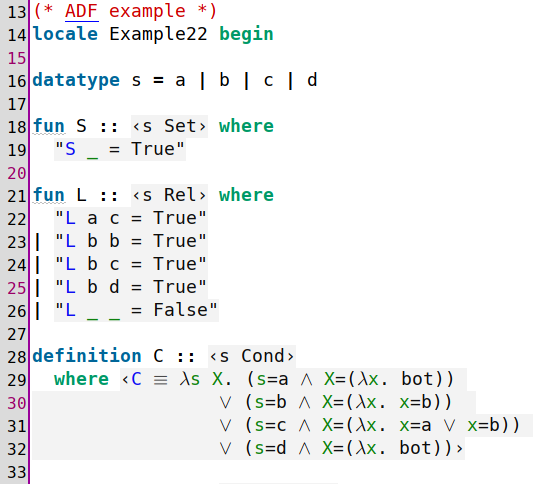}
\end{minipage}\;
\begin{minipage}{.5\textwidth}
  \includegraphics[height=5cm,interpolate]{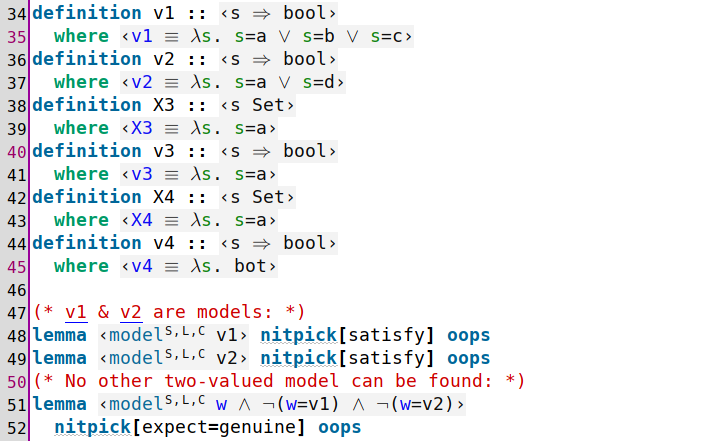}
\end{minipage}
\caption{Isabelle/HOL source file after encoding the ADF example from Fig.~\ref{fig:exampleADF}.}
\label{fig:isabelle}
\end{figure}

In order to encode the ADF example from Fig.~\ref{fig:exampleADF} in Isabelle/HOL, (a) a type for the arguments involved, (b) the different components of the ADF, and (c) the different interpretations are encoded. Their representation in Isabelle/HOL is displayed in Fig.~\ref{fig:isabelle}.

A new datatype $s$ is defined in Isabelle/HOL that represents the arguments of the example. Following the four arguments in the example, the domain associated with type $s$ is defined to be $D_s = \{a, b, c, d\}$.

The corresponding ADF $(S,L,C)$ is then encoded in HOL as follows:
\begin{itemize}
  \item $S := \lambda s.\ \top$
  \item $\begin{aligned}[t]
    L := \lambda s1.\, \lambda s2.\, & (s1=a \land s2=c) \lor (s1=b \land s2=b) \; \lor \\
    &(s1=b \land s2=c) \lor (s1=b \land s2=d)
\end{aligned}$
  \item $\begin{aligned}[t]
    C := \lambda s.\,\lambda X.\, & (s=a \land X= (\lambda x.\, \bot)) \; \lor &&\\
    &(s=b \land X=(\lambda x.\, x=b)) \;\lor &&\\
    &(s=c \land X=(\lambda x.\, x=a \lor x=b)) \;\lor &&\\
    &(s=d \land X=(\lambda x.\, \bot))
\end{aligned}$
\end{itemize}

Next, the interpretations $v_1 = \{a \mapsto  \textbf{t}, b \mapsto \textbf{t}, c \mapsto \textbf{t}, d \mapsto \textbf{f}\}$, $v_2 = \{a \mapsto  \textbf{t}, b \mapsto \textbf{f}, c \mapsto \textbf{f}, d \mapsto \textbf{t}\}$, $v_3 = \{a \mapsto  \textbf{t}, b \mapsto \textbf{u}, c \mapsto \textbf{u}, d \mapsto \textbf{u}\}$, and $v_4 = \{a \mapsto  \textbf{f}, b \mapsto \textbf{u}, c \mapsto \textbf{u}, d \mapsto \textbf{u}\}$ are
encoded as follows:

\begin{itemize}
  \item $v1 := \lambda s.\, s=a \lor s=b \lor s=c$
  \item $v2 := \lambda s.\, s=a \lor s=d$
  \item $v3 := \lambda s.\, s=a$
  \item $v4 := \lambda s.\, \bot$
  \item $X3 := \lambda s.\, s=a$
  \item $X4 := \lambda s.\, s=a$
\end{itemize}

Finally, it can be established that the above encoded interpretations behave as expected:
\begin{itemize} 
\item $v1$ and $v2$ are models, i.e., $\texttt{model}^{S,L,C}\ v1$ and $\texttt{model}^{S,L,C}\ v2$.
\item No other two-valued model can be found: No model can be found for 
$(\texttt{model}^{S,L,C}\ w) \land \neg(w=v1) \land \neg(w=v2)$
\item $v1$, $v2$ and $v3$ are admissible:\\
$\texttt{admissible}^{S,L,C}\ S\ v1$, 
$\texttt{admissible}^{S,L,C}\ S\ v2$,
$\texttt{admissible}^{S,L,C}\ X3\ v3$
\item $v4$ is not admissible: $\texttt{admissible}^{S,L,C}\ X4\ v4$ has no model.
\item $v1$, $v2$ and $v3$ are complete:\\
$\texttt{complete}^{S,L,C}\ S\ v1$,
$\texttt{complete}^{S,L,C}\ S\ v2$,
$\texttt{complete}^{S,L,C}\ X3\ v3$.
\item No other interpretation is complete: The expression \ldots \\
$(\texttt{complete}^{S,L,C}\ X\ v) \land\\ \big(\neg(X=S) \lor \neg(v=v1)\big) \land \big(\neg(X=S) \lor \neg(v=v2)\big) \land \big(\neg(X=X3) \lor \neg(v=v3)\big)$ has no model.
\item $v1$ and $v2$ are preferred: $\texttt{preferred}^{S,L,C}\ S\ v1$, $\texttt{preferred}^{S,L,C}\ S\ v2$.
\item No other interpretation is preferred: The expression \ldots \\
$(\texttt{preferred}^{S,L,C}\ X\ v) \land \big(\neg(X=S) \lor \neg(v=v1)\big) \land \big(\neg(X=S) \lor \neg(v=v2)\big)$ has no model.
\item $v3$ is the grounded interpretation: $\texttt{grounded}^{S,L,C}\ X3\ v3$.
\item No other interpretation is grounded: The expression \ldots \\
$(\texttt{grounded}^{S,L,C}\ X\ v) \land \big(\neg(X=X3) \lor \neg(v=v3)\big)$ has no model.
\item $v2$ is a stable model: $\texttt{stable}^{S,L,C}\ v2$.
\item $v1$ is not a stable model: $\texttt{stable}^{S,L,C}\ v1$ has no model.
\item No other interpretation is stable: $(\texttt{stable}^{S,L,C}\ w) \land \neg(w=v2)$ has no model.
\end{itemize}

As can be seen, Isabelle/HOL can be seen to conduct interactive experiments with given ADFs.

\subsubsection{Example AF.}
\begin{figure}[tb]
    \centering
    \includegraphics[scale=0.4]{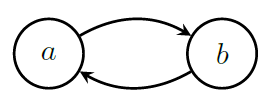}
    \caption{An example AF.}
    \label{fig:AF}
\end{figure}

In the second example a simple AF, as visualized by Fig.~\ref{fig:AF}, is assessed using Isabelle/HOL.

Analogously to the previous example, a new datatype $s$ is introduced to represent the type of arguments present in the AF; it is defined such that the domain associated with type $s$ is $D_s = \{a, b\}$ (two arguments). The AF $(S,L,C)$ is then encoded as follows:
\begin{itemize}
  \item $S := \lambda s.\ \top$
  \item $L := \lambda s1.\, \lambda s2.\, (s1=a \land s2=b) \lor (s1=b \land s2=a)$
  \item $C := \lambda s.\, \lambda X.\, (s=a \land X=(\lambda x.\, \bot)) \lor (s=b \land X=(\lambda x.\, \bot))$
\end{itemize}

The interpretations $v_1 = \{a \mapsto \textbf{t}, b \mapsto \textbf{f}\}$, $v_2 = \{a \mapsto \textbf{f}, b \mapsto \textbf{t}\}$, $v_3 = \{a \mapsto \textbf{u}, b \mapsto \textbf{u}\}$, $v_4 = \{a \mapsto \textbf{t}, b \mapsto \textbf{t}\}$, and $v_5 = \{a \mapsto \textbf{f}, b \mapsto \textbf{f}\}$ are then encoded as:
\begin{itemize}
  \item $v1 := \lambda s.\, s=a$
  \item $v2 := \lambda s.\, s=b$
  \item $v3 := \lambda s.\, \bot$
  \item $v4 := \lambda s.\, s=a \lor s=b$
  \item $v5 := \lambda s.\, \bot$
  \item $X3 := \lambda s.\, \bot$
\end{itemize}

Again, the encoded interpretations behave as expected as verified within Isabelle/HOL:
\begin{itemize} 
\item $v1$ and $v2$ are models: $\texttt{model}^{S,L,C}\ v1$, $\texttt{model}^{S,L,C}\ v2$.
\item No other two-valued model can be found: The expression 
$(\texttt{model}^{S,L,C}\ w) \land \neg(w=v1) \land \neg(w=v2)$ has no model.
\item $v1$ and $v2$ are stable: $\texttt{stable}^{S,L,C}\ v1$, $\texttt{stable}^{S,L,C}\ v2$.
\item No other stable interpretation can be found: The expression
$(\texttt{stable}^{S,L,C}\ w) \land \neg(w=v1) \land \neg(w=v2)$ has no model.
\item $v1$, $v2$ and $v3$ are admissible: $\texttt{admissible}^{S,L,C}\ S\ v1$, $\texttt{admissible}^{S,L,C}\ S\ v2$, 
$\texttt{admissible}^{S,L,C}\ X3\ v3$.
\item $v4$ and $v5$ is not admissible: Both $\texttt{admissible}^{S,L,C}\ S\ v4$ and $\texttt{admissible}^{S,L,C}\ S\ v5$
 have no model.
\item $v1$, $v2$ and $v3$ are complete: $\texttt{complete}^{S,L,C}\ S\ v1$, 
$\texttt{complete}^{S,L,C}\ S\ v2$, $\texttt{complete}^{S,L,C}\ X3\ v3$.
\item No other interpretation is complete (in particular $v4$ and $v5$ are not complete): The expression 
$(\texttt{complete}^{S,L,C}\ X\ v) \land\\ \big(\neg(X=S) \lor \neg(v=v1)\big) \land \big(\neg(X=S) \lor \neg(v=v2)\big) \land \big(\neg(X=X3) \lor \neg(v=v3)\big)$ has no model.
\item $v1$ and $v2$ are preferred: $\texttt{preferred}^{S,L,C}\ S\ v1$, $\texttt{preferred}^{S,L,C}\ S\ v2$.
\item No other interpretation is preferred: The expression 
$(\texttt{preferred}^{S,L,C}\ X\ v) \land \big(\neg(X=S) \lor \neg(v=v1)\big) \land \big(\neg(X=S) \lor \neg(v=v2)\big)$ has no model.
\item $v3$ is the grounded interpretation: $\texttt{grounded}^{S,L,C}\ X3\ v3$.
\item No other interpretation is grounded: The expression \\
$(\texttt{grounded}^{S,L,C}\ X\ v) \land \big(\neg(X=X3) \lor \neg(v=v3)\big)$ has no model.
\end{itemize}

The presented encoding thus can represent and assess both AFs and ADFs within one logical framework.

\section{Conclusion}
In this article, an encoding of abstract dialectical frameworks (ADFs) \textbf{--} a generalisation of Dung's argumentation frameworks (AFs) \textbf{--} and of several standard semantics found in literature into higher-order logic was presented. In order to do so, sets are represented by their characteristic function, and three-valued interpretations are encoded as a pair of sets: one set containing the statements mapped to either $True$ or $False$ by the interpretation, and another set containing the statements mapped to $True$ only. Finally, argumentation semantics were encoded as higher-order predicates that, by construction, return $True$ if and only if the input interpretation fulfils the definition of said semantics.

The encoding was implemented in Isabelle/HOL, a well-known HOL-based proof assistant. This choice was motivated by the many practical features of Isabelle/HOL, in particular the Sledgehammer system and the model-finder Nitpick. The encoding's source files (so-called \emph{theory files} in the context of Isabelle/HOL) are freely available at Zenodo \cite{files}. It should be noted that the encoding presented in this article is not limited to any specific reasoning system (as long as it at least as expressive as HOL); and the focus on Isabelle/HOL has been simply for demonstration purposes because of its flexibility and user-friendliness, such as, for example, a graphical user interface for interactive experimentation and its portfolio of integrated automated reasoning tools.

The expressiveness of HOL allows to perform both meta-level reasoning on ADFs (reasoning \textit{about} ADFs) as well as object-level reasoning (reasoning \textit{with} ADFs). Regarding meta-level reasoning, several fundamental and well-known properties of the operator-based ADF formalism used as basis for the encoding were formally proven. Furthermore, well-known relations between several standard semantics found in literature were also formally encoded and proven. Additionally, properties that should not hold were disproved by counter-models generated by Nitpick. Concerning object-level reasoning, two application examples were examined. 

It should be noted that the approach described in this work is not necessarily meant to provide an alternative to efficient and well-established methods used for computing interpretations of large-scale AFs or ADFs. The goal is rather to bridge the gap between ADFs and automated reasoning in order to provide generic means to automatically and interactively assess ADFs within a rich ecosystem of reasoning tools. In fact, this work is line with the idea behind the LogiKEy framework \cite{LogiKEy}, where classic HOL is used for the semantical embedding of normative theories for ethical and legal reasoning.

\paragraph{Further work.}

This work could be extended by verifying further properties of the ultimate approxiations and the encoded ADF semantics. Also the encoding of new semantics proposed recently in the literature provides a relevant research direction, and performing further meta-theoretical studies on such semantics. Some examples of new ADF semantics include $nai_2$ and $stg_2$ \cite{new_ADF_semantics}, which are the ADF counterparts to the AF semantics $cf2$ \cite{cf2_AF_semantics} and $stage2$ \cite{stage_AF_semantics}.

Additionally, due to the expressiveness of HOL, other variations, extensions or sub-classes of ADFs could be encoded, explored and further investigated. For example, a rather expressive class of ADFs that is not only of practical interest but also has nice computational properties is the class of bipolar ADFs \cite{ADFs,bipolar_ADFs}. Some other interesting extensions that may be further studied using our approach include, for example, weighted \cite{ADFs,weighted_ADFs,weighted_ADFs_AFT} and preference-based \cite{ADFs,preferences_ADFs} ADFs.

\section*{Acknowledgements}
This work was supported by the Luxembourg National Research Fund (FNR) under grant C20/IS/14616644 (AuReLeE).


\bibliographystyle{plain}
\bibliography{main}



\end{document}

%% file: commands.tex
\newcommand{\B}{\mathcal{B}}

\newcommand{\V}{\mathcal{V}}
\newcommand{\T}{\mathcal{T}}

\newcommand{\bo}{o}
\newcommand{\ar}{\Rightarrow}



%% file: main.bbl
\begin{thebibliography}{10}

\bibitem{amgoud2010handling}
Leila Amgoud and Srdjan Vesic.
\newblock {Handling Inconsistency with Preference-Based Argumentation}.
\newblock In {\em International Conference on Scalable Uncertainty Management},
  pages 56--69, 2010.

\bibitem{arieli2013qbf}
Ofer Arieli and Martin~WA Caminada.
\newblock {A QBF-based formalization of abstract argumentation semantics}.
\newblock {\em Journal of Applied Logic}, 11(2):229--252, 2013.

\bibitem{AFsHandbook}
Pietro Baroni, Martin Caminada, and Massimiliano Giacomin.
\newblock Abstract {A}rgumentation {F}rameworks and {T}heir {S}emantics.
\newblock In Pietro Baroni, Dov~M. Gabbay, Massimiliano Giacomin, and Leendert
  van~der Torre, editors, {\em Handbook of Formal Argumentation}, pages
  159--236. College Publications, 2018.

\bibitem{cf2_AF_semantics}
Pietro Baroni, Massimiliano Giacomin, and Giovanni Guida.
\newblock {SCC-recursiveness: A general schema for argumentation semantics}.
\newblock {\em Artificial Intelligence}, 168(1-2):162--210, 2005.

\bibitem{baroni2020acceptability}
Pietro Baroni, Francesca Toni, and Bart Verheij.
\newblock On the acceptability of arguments and its fundamental role in
  nonmonotonic reasoning, logic programming and n-person games: 25 years later.
\newblock {\em Argument Comput.}, 11(1-2):1--14, 2020.

\bibitem{CVC4-solver}
Clark Barrett, Christopher~L Conway, Morgan Deters, Liana Hadarean, Dejan
  Jovanovi{\'c}, Tim King, Andrew Reynolds, and Cesare Tinelli.
\newblock {CVC4}.
\newblock In {\em International Conference on Computer Aided Verification},
  pages 171--177, 2011.

\bibitem{beierle2015software}
Christoph Beierle, Florian Brons, and Nico Potyka.
\newblock {A software system using a SAT solver for reasoning under complete,
  stable, preferred, and grounded argumentation semantics}.
\newblock In {\em Joint German/Austrian Conference on Artificial Intelligence
  (K{\"u}nstliche Intelligenz)}, pages 241--248, 2015.

\bibitem{Currying}
Christoph Benzm{\"u}ller and Peter Andrews.
\newblock Church's {T}ype {T}heory.
\newblock {\em Stanford Encyclopedia of Philosophy}, pages 1--62, 2019.

\bibitem{automationOfHOL}
Christoph Benzm{\"u}ller and Dale Miller.
\newblock Automation of higher-order logic.
\newblock In Dov~M. Gabbay, J\"org~H. Siekmann, and John Woods, editors, {\em
  Handbook of the History of Logic, Volume 9 --- Computational Logic}, pages
  215--254. North Holland, Elsevier, 2014.

\bibitem{LogiKEy}
Christoph Benzm{\"u}ller, Xavier Parent, and Leendert van~der Torre.
\newblock Designing normative theories for ethical and legal reasoning:
  {LogiKEy} framework, methodology, and tool support.
\newblock {\em Artificial intelligence}, 287:103348, 2020.

\bibitem{besnard2004checking}
Philippe Besnard and Sylvie Doutre.
\newblock Checking the acceptability of a set of arguments.
\newblock In {\em Proceedings of the 10th International Workshop on
  Non-Monotonic Reasoning ({NMR} 2004)}, pages 59--64, 2004.

\bibitem{biere2009handbook}
Armin Biere, Marijn Heule, Hans van Maaren, and Toby Walsh, editors.
\newblock {\em Handbook of Satisfiability}, volume 185 of {\em Frontiers in
  Artificial Intelligence and Applications}. {IOS} Press, 2009.

\bibitem{Sledgehammer}
Jasmin~Christian Blanchette, Sascha B{\"o}hme, and Lawrence~C Paulson.
\newblock Extending {S}ledgehammer with {SMT} {S}olvers.
\newblock {\em Journal of automated reasoning}, 51(1):109--128, 2013.

\bibitem{Nitpick}
Jasmin~Christian Blanchette and Tobias Nipkow.
\newblock Nitpick: A {C}ounterexample {G}enerator for {H}igher-{O}rder {L}ogic
  {B}ased on a {R}elational {M}odel {F}inder.
\newblock In {\em International Conference on Interactive Theorem Proving},
  pages 131--146, 2010.

\bibitem{weighted_ADFs_AFT}
Bart Bogaerts.
\newblock {Weighted Abstract Dialectical Frameworks through the Lens of
  Approximation Fixpoint Theory}.
\newblock {\em Proceedings of the AAAI Conference on Artificial Intelligence},
  33(01):2686--2693, 2019.

\bibitem{ADFs}
Gerhard Brewka, Stefan Ellmauthaler, Hannes Strass, Johannes~P. Wallner, and
  Stefan Woltran.
\newblock Abstract dialectical frameworks.
\newblock In Pietro Baroni, Dov~M. Gabbay, Massimiliano Giacomin, and Leendert
  van~der Torre, editors, {\em Handbook of formal argumentation}, pages
  237--285. College Publications, 2018.

\bibitem{Brewka2013}
Gerhard Brewka, Stefan Ellmauthaler, Hannes Strass, Johannes~Peter Wallner, and
  Stefan Woltran.
\newblock Abstract dialectical frameworks revisited.
\newblock In {\em Proceedings of the Twenty-Third international joint
  conference on Artificial Intelligence}, pages 803--809, 2013.

\bibitem{preferences_ADFs}
Gerhard Brewka, Stefan Ellmauthaler, Hannes Strass, Johannes~Peter Wallner, and
  Stefan Woltran.
\newblock {Abstract Dialectical Frameworks Revisited}.
\newblock In {\em Proceedings of the Twenty-Third international joint
  conference on Artificial Intelligence}, pages 803--809, 2013.

\bibitem{weighted_ADFs}
Gerhard Brewka, Hannes Strass, Johannes Wallner, and Stefan Woltran.
\newblock {Weighted Abstract Dialectical Frameworks}.
\newblock {\em Proceedings of the AAAI Conference on Artificial Intelligence},
  32(1), 2018.

\bibitem{bipolar_ADFs}
Gerhard Brewka and Stefan Woltran.
\newblock {Abstract Dialectical Frameworks}.
\newblock In {\em Proceedings of the Twelfth International Conference on
  Principles of Knowledge Representation and Reasoning}, pages 102--111, 2010.

\bibitem{cayrol2020logical}
Claudette Cayrol and Marie-Christine Lagasquie-Schiex.
\newblock {Logical Encoding of Argumentation Frameworks with Higher-order
  Attacks and Evidential Supports}.
\newblock {\em International Journal on Artificial Intelligence Tools},
  29(03n04):2060003:1--2060003:50, 2020.

\bibitem{cerutti2017foundations}
Federico Cerutti, Sarah~A Gaggl, Matthias Thimm, and Johannes Wallner.
\newblock Foundations of implementations for formal argumentation.
\newblock {\em IfCoLog Journal of Logics and their Applications},
  4(8):2623--2705, 2017.

\bibitem{cerutti2017efficient}
Federico Cerutti, Mauro Vallati, and Massimiliano Giacomin.
\newblock {An Efficient Java-Based Solver for Abstract Argumentation
  Frameworks: jArgSemSAT}.
\newblock {\em International Journal on Artificial Intelligence Tools},
  26(02):1750002:1–--1750002:26, 2017.

\bibitem{Church}
Alonzo Church.
\newblock A formulation of the simple theory of types.
\newblock {\em Journal of Symbolic Logic}, 5(2):56--68, 1940.

\bibitem{de2016argumentation}
Florence~Dupin de~Saint-Cyr, Pierre Bisquert, Claudette Cayrol, and
  Marie-Christine Lagasquie-Schiex.
\newblock {Argumentation update in YALLA (Yet Another Logic Language for
  Argumentation)}.
\newblock {\em International Journal of Approximate Reasoning}, 75:57--92,
  2016.

\bibitem{DMT2000}
Marc Denecker, Victor~W Marek, and Miros{\l}aw Truszczy{\'n}ski.
\newblock Approximations, {S}table {O}perators, {W}ell-{F}ounded {F}ixpoints
  and {A}pplications in {N}onmonotonic {R}easoning.
\newblock In {\em Logic-Based Artificial Intelligence}, pages 127--144, 2000.

\bibitem{DMT2004}
Marc Denecker, Victor~W Marek, and Miros{\l}aw Truszczy{\'n}ski.
\newblock Ultimate approximation and its application in nonmonotonic knowledge
  representation systems.
\newblock {\em Information and Computation}, 192(1):84--121, 2004.

\bibitem{DBLP:journals/argcom/DillerWW15}
Martin Diller, Johannes~Peter Wallner, and Stefan Woltran.
\newblock Reasoning in abstract dialectical frameworks using quantified boolean
  formulas.
\newblock {\em Argument Comput.}, 6(2):149--177, 2015.

\bibitem{AFsDung}
Phan~Minh Dung.
\newblock On the acceptability of arguments and its fundamental role in
  nonmonotonic reasoning, logic programming and n-person games.
\newblock {\em Artificial intelligence}, 77(2):321--357, 1995.

\bibitem{dunne2002coherence}
Paul~E Dunne and Trevor~JM Bench-Capon.
\newblock Coherence in finite argument systems.
\newblock {\em Artificial Intelligence}, 141(1-2):187--203, 2002.

\bibitem{stage_AF_semantics}
Wolfgang Dvo{\v{r}}{\'a}k and Sarah~Alice Gaggl.
\newblock {Stage semantics and the SCC-recursive schema for argumentation
  semantics}.
\newblock {\em Journal of Logic and Computation}, 26(4):1149--1202, 2014.

\bibitem{dvovrak2014complexity}
Wolfgang Dvo{\v{r}}{\'a}k, Matti J{\"a}rvisalo, Johannes~Peter Wallner, and
  Stefan Woltran.
\newblock Complexity-sensitive decision procedures for abstract argumentation.
\newblock {\em Artificial Intelligence}, 206:53--78, 2014.

\bibitem{egly2006reasoning}
Uwe Egly and Stefan Woltran.
\newblock {Reasoning in Argumentation Frameworks Using Quantified Boolean
  Formulas}.
\newblock {\em Computational Models of Argument: Proceedings of COMMA},
  6:133--144, 2006.

\bibitem{DBLP:conf/comma/EllmauthalerS14}
Stefan Ellmauthaler and Hannes Strass.
\newblock The {DIAMOND} system for computing with abstract dialectical
  frameworks.
\newblock In {\em {COMMA}}, volume 266 of {\em Frontiers in Artificial
  Intelligence and Applications}, pages 233--240. {IOS} Press, 2014.

\bibitem{gabbay2013}
Dov~M Gabbay.
\newblock {\em {Meta-logical Investigations in Argumentation Networks}},
  volume~44 of {\em Studies in Logic – Mathematical Logic and Foundations}.
\newblock College Publications, 2013.

\bibitem{new_ADF_semantics}
Sarah~Alice Gaggl, Sebastian Rudolph, and Hannes Strass.
\newblock {On the Decomposition of Abstract Dialectical Frameworks and the
  Complexity of Naive-based Semantics}.
\newblock {\em Journal of Artificial Intelligence Research}, 70:1--64, 2021.

\bibitem{Henkin1}
Leon Henkin.
\newblock Completeness in the theory of types.
\newblock {\em Journal of Symbolic Logic}, 15(2):81--91, 1950.

\bibitem{DBLP:journals/ai/LinsbichlerMNWW22}
Thomas Linsbichler, Marco Maratea, Andreas Niskanen, Johannes~Peter Wallner,
  and Stefan Woltran.
\newblock Advanced algorithms for abstract dialectical frameworks based on
  complexity analysis of subclasses and {SAT} solving.
\newblock {\em Artif. Intell.}, 307:103697, 2022.

\bibitem{files}
Antoine Martina.
\newblock {An Encoding of Abstract Dialectical Frameworks into Higher-Order
  Logic}.
\newblock Supplemental material to the article: Isabelle/HOL sources files.

\bibitem{Z3-solver}
Leonardo~de Moura and Nikolaj Bj{\o}rner.
\newblock Z3: {A}n {E}fficient {SMT} {S}olver.
\newblock In {\em International Conference on Tools and Algorithms for the
  Construction and Analysis of Systems}, pages 337--340, 2008.

\bibitem{Isabelle/HOL}
Tobias Nipkow, Markus Wenzel, and Lawrence~C Paulson.
\newblock {\em Isabelle/{HOL}: {A} {P}roof {A}ssistant for {H}igher-{O}rder
  {L}ogic}.
\newblock Lecture Notes in Computer Science 2283. Springer, 2002.

\bibitem{E-prover}
Stephan Schulz.
\newblock E -- a brainiac theorem prover.
\newblock {\em AI Communications}, 15(2-3):111--126, 2002.

\bibitem{rahwan2009argumentation}
Guillermo~Ricardo Simari and Iyad Rahwan, editors.
\newblock {\em Argumentation in Artificial Intelligence}.
\newblock Springer, 2009.

\bibitem{AlexPhD}
Alexander Steen.
\newblock Extensional {P}aramodulation for {H}igher-{O}rder {L}ogic and its
  {E}ffective {I}mplementation {L}eo-{III}.
\newblock {\em KI-K{\"u}nstliche Intelligenz}, 34(1):105--108, 2020.

\bibitem{SB2021}
Alexander Steen and Christoph Benzm{\"{u}}ller.
\newblock {Extensional Higher-Order Paramodulation in Leo-III}.
\newblock {\em Journal of Automated Reasoning}, 65(6):775--807, 2021.

\bibitem{AFinHOL}
Alexander Steen and David Fuenmayor.
\newblock A {F}ormalisation of {A}bstract {A}rgumentation in {H}igher-{O}rder
  {L}ogic.
\newblock {\em Journal of Logic and Computation}, 09 2023.
\newblock exac027.

\bibitem{Strass2013a}
Hannes Strass.
\newblock Approximating operators and semantics for abstract dialectical
  frameworks.
\newblock {\em Artificial Intelligence}, 205:39--70, 2013.

\bibitem{Strass&Wallner2015}
Hannes Strass and Johannes~Peter Wallner.
\newblock Analyzing the computational complexity of abstract dialectical
  frameworks via approximation fixpoint theory.
\newblock {\em Artificial Intelligence}, 226:34--74, 2015.

\bibitem{toni2011argumentation}
Francesca Toni and Marek Sergot.
\newblock {Argumentation and Answer Set Programming}.
\newblock {\em Logic Programming, Knowledge Representation, and Nonmonotonic
  Reasoning}, pages 164--180, 2011.

\bibitem{wallner2013advanced}
Johannes~Peter Wallner, Georg Weissenbacher, and Stefan Woltran.
\newblock {Advanced SAT techniques for abstract argumentation}.
\newblock In {\em International Workshop on Computational Logic in Multi-Agent
  Systems}, pages 138--154, 2013.

\bibitem{DBLP:conf/tphol/WenzelP06}
Markus Wenzel and Lawrence~C. Paulson.
\newblock Isabelle/isar.
\newblock In {\em The Seventeen Provers of the World}, volume 3600 of {\em
  Lecture Notes in Computer Science}, pages 41--49. Springer, 2006.

\bibitem{zhang2009argumentation}
Xiaowang Zhang and Zuoquan Lin.
\newblock {An Argumentation-Based Approach to Handling Inconsistencies in
  DL-Lite}.
\newblock In {\em Annual Conference on Artificial Intelligence}, pages
  615--622, 2009.

\end{thebibliography}
